\documentclass[usenatbib,fleqn]{mn2e}

\usepackage[T1]{fontenc}
\usepackage[utf8]{inputenc}

\usepackage{pdflscape}
\usepackage{physics}
\usepackage{textcomp}
\usepackage{bm}
\usepackage{comment}
\usepackage{multicol}
\usepackage{multirow}
\usepackage{booktabs}
\usepackage{times,amsmath,amssymb,longtable,breqn}
\usepackage[varg]{txfonts}
\usepackage{lastpage}

\usepackage{grffile} 
\usepackage{graphicx}

\usepackage{color}
\usepackage[dvipsnames]{xcolor}
\usepackage{soul}

\newcommand\RB[1]{{\color{DarkOrchid}{#1}}}

\usepackage{ulem}
\newcommand\U[1]{{\,\rm #1}}

\newcommand\al{\alpha}
\newcommand\bt{\beta}

\newcommand\dl{\delta}
\newcommand\lmb{\lambda}
\newcommand\om{\omega}
\newcommand\Msun{M_\odot}
\newcommand\rs[1]{_\mathrm{#1}}

\newcommand\Esn{E\rs{sn}}
\newcommand\Mej{M\rs{ej}}
\newcommand\rhoej{\rho\rs{ej}}
\newcommand\rhoism{\rho\rs{0}}
\newcommand\nism{n\rs{0}}

\newcommand\Rch{R\rs{ch}}
\newcommand\tch{t\rs{ch}}
\newcommand\Vch{V\rs{ch}}

\newcommand\vt{v\rs{t}}

\newcommand\Rrev{R\rs{RS}}
\newcommand\Rcont{R\rs{CD}}
\newcommand\Rforw{R\rs{FS}}

\newcommand\tini{t\rs{ini}}
\newcommand\tcore{t\rs{core}}
\newcommand\tcoreCD{t\rs{core,CD}}
\newcommand\tcoreFS{t\rs{core,FS}}

\newcommand\timplo{t\rs{implo}}

\newcommand\btimplo{\bt\rs{i}}

\newcommand\epsRS{\epsilon\rs{RS}}
\newcommand\aRS{a\rs{RS}}
\newcommand\bRS{b\rs{RS}}
\newcommand\cRS{c\rs{RS}}

\newcommand\aCD{a\rs{CD}}
\newcommand\bCD{b\rs{CD}}
\newcommand\cCD{c\rs{CD}}

\begin{document}
\label{firstpage}
\pagerange{\pageref{firstpage}--\pageref{lastpage}}

\title[Revisiting the evolution of nonradiative SNRs]{Revisiting the evolution of nonradiative supernova remnants: \\ 
 A hydrodynamical-informed parameterization of the shock positions }
\author[Bandiera et al.]
{R. Bandiera$^{1}$, N. Bucciantini$^{1,2,3}$, J. Mart\'in$^{4,5}$, 
B. Olmi$^{1,4,7}$, D. F. Torres$^{4,5,6}$ \thanks{All authors have contributed equally to this work.}  \\
$^{1}$ INAF - Osservatorio Astrofisico di Arcetri, Largo E. Fermi 5, I-50125 Firenze, Italy \\
$^{2}$ Dipartamento de Fisica e Astronomia, Universit\`a degli Studi di Firenze, Via G. Sansone 1, I-50019 Sesto F. no (Firenze), Italy \\
$^{3}$ INFN - Sezione di Firenze, Via G. Sansone 1, I-50019 Sesto F. no (Firenze), Italy \\
$^{4}$ Institute of Space Sciences (ICE, CSIC), Campus UAB, Carrer de Can Magrans s/n, 08193 Barcelona, Spain \\
$^{5}$ Institut d'Estudis Espacials de Catalunya (IEEC), Gran Capit\`a 2-4, 08034 Barcelona, Spain \\
$^{6}$ Instituci\'o Catalana de Recerca i Estudis Avan\c cats (ICREA), 08010 Barcelona, Spain \\
$^{7}$ INAF - Osservatorio Astronomico di Palermo, Piazza del Parlamento 1, I-90134 Palermo, Italy
}

\date{}
\maketitle

\pubyear{2021}

\begin{abstract}
Understanding   
the evolution of a supernova remnant shell in time is fundamental.
 Such understanding is critical to build reliable models of 
the dynamics of the supernova remnant shell  interaction with any pulsar wind nebula it might contain.
Here, we perform a large study of the parameter space for the one-dimensional spherically symmetric evolution of a supernova remnant,  accompanying it by analytical analysis.
Assuming, as is usual, an ejecta density profile with a power-law core and an envelope, and a uniform ambient medium, we provide a set of highly-accurate approximations for the evolution of the main structural features of supernova remnants, such as the reverse and forward shocks and the contact discontinuity. 
We compare our results with previously adopted approximations, showing that  existing simplified prescriptions can easily lead to large errors.
In particular, in the context of pulsar wind nebulae modelling, an accurate description for the supernova remnant reverse shock is required.
We also study in depth the self-similar solutions for the  initial phase of evolution, when the reverse shock propagates through the  envelope  of  the  ejecta.  
Since  these  self-similar  solutions  are exact, but not fully analytical, we here provide highly-accurate approximations as well. 
\end{abstract}



\section{Introduction}
\label{sec:intro}
Neglecting instabilities, clumpiness, and gradients in the ambient medium, the evolution of a SNR within 
it can be described in a schematic way in terms of a shell bounded by two shocks: the expanding SN ejecta drive a forward shock (FS) in the ambient medium and a reverse shock (RS), that moves through the ejecta.
In addition, a contact discontinuity (CD) separates the ejecta material and the shocked ambient medium.

A detailed description of the nonradiative evolution of SNRs was presented  by \citet[hereafter TM99]{Truelove1999}.
In that work, through a mix of analytic limits, semi-analytical formulae, and fits to numerical simulations, a series of approximations to describe the evolution of the SNR during its different stages was provided.
This paper has become a widely used reference for the time evolution of the FS and RS.

 TM99 solutions are used, for instance, when incorporating the dynamics in radiative models of PWNe, see, e.g., \cite{Gelfand_Slane+09a,Bucciantini_Arons+11a,Martin2012,Torres:2014}.
In \citet{Bandiera:2020}, we have used these solutions as well for modelling the physical conditions at the beginning of the so called \textit{reverberation} phase.
This time is identified as the moment in which the boundary of the PWN, formerly expanding into the unshocked ejecta, reaches the RS.
These initial conditions are a key ingredient for modelling the late PWN evolution, and ultimately to estimate the PWN compression, but they are highly sensitive to the RS properties (position and velocity) immediately before the beginning of the reverberation phase.
This in particular will be the argument of a forthcoming paper in our reverberation project, started with \citet{Bandiera:2020}.

Several attempts to reproduce and  improve the TM99 model were made in the past, considering different parametrizations for the ambient medium \citep{Tang2017} or a complex clumpy structure for the ejecta \citep{Micelotta2016}. 
Nevertheless these works focus on specific problems and/or objects, and a more general description of the shocks (and CD) evolution with varying the characteristic parameters of the problem is still not available in the literature.
To this purpose we have run a large number of numerical simulations, which span a wide range of power-law indices for the envelope of the ejecta,
as well as a choice of density slopes for their core. Combining these numerical data with known results of self-similar models, valid during the early SNR expansion \cite[based on][C82 hereafter]{Chevalier1982}, we are able to derive highly-accurate prescriptions covering a wide region of the parameter space of possible ejecta structures, showing that there are general trends that allow one to re-scale the evolution of the main structural features.

This paper is organized as follows.
In Sec.~\ref{sec:basic} we recall the definitions of the characteristic physical scales of the problem and the typical assumptions for the ejecta profiles.
In Sec.~\ref{sec:selfsim} we discuss the early evolution of the SNR, which is fully described by the self-similar models of C82.  We improve on this description there as well. 
In Sec.~\ref{sec:numerical} we describe the scheme used for our numerical simulations.
In Sec.~\ref{sec:snr} we present the results of our numerical models to reproduce the evolution of SNRs with varying the parameters that define the ejecta morphology (namely the core and envelope structures).
We discuss our findings and compare with TM99 results,  focusing on the limitations of that work when trying to reconstruct the evolution with a high accuracy.
In Sec.~\ref{sec:traj} we present our new set of analytic approximations for the RS, CD and FS time evolution, discussing their validity and precision.
Finally our conclusions are drawn in Sec.~\ref{sec:concl}.
%


\section{Basic assumptions}
\label{sec:basic}
As shown by TM99, the SNR evolution can be naturally scaled in terms of the characteristic length $\Rch$ and time $\tch$ given by:
\begin{eqnarray}
\Rch\!\!\!&=&\!\!\!
\Mej^{1/3}\rhoism^{-1/3}		\nonumber\\
\label{eq:Rchdef}
\!\!\!&\simeq&\!\!\!
7.4\U{pc}\left(\frac{\Mej}{10\,\Msun}\right)^{1/3}\left(\frac{m_p\nism}{\U{g\, cm^{-3}}}\right)^{-1/3}\,,			\\
\tch\!\!\!&=&\!\!\!
\Esn^{-1/2}\Mej^{5/6}\rhoism^{-1/3}	\nonumber\\
\label{eq:tchdef}
\!\!\!&\simeq&\!\!\!
3241\U{yr}\left(\frac{\Esn}{10^{51}\U{erg}}\right)^{-1/2}\left(\frac{\Mej}{10\,\Msun}\right)^{5/6}\left(\frac{m_p\nism}{\U{g\, cm^{-3}}}\right)^{-1/3}\,,
\end{eqnarray}
where $\Mej$ is the ejecta mass,  $\Esn$ the SN energy, $\rhoism$ the ambient medium density and $\nism$ the related number density, while $m_p$ is the proton mass.
From these one can also define a velocity scale as:
\begin{equation}
\!\Vch\,=\,\frac{\Esn^{1/2}}{\Mej^{1/2}}\,\simeq\,2240\U{km\,s^{-1}}\left(\frac{\Esn}{10^{51}\U{erg}}\right)^{1/2}\left(\frac{\Mej}{10\,\Msun}\right)^{-1/2}\,.
\end{equation}
\begin{figure}
\centering
	\includegraphics[width=.47\textwidth]{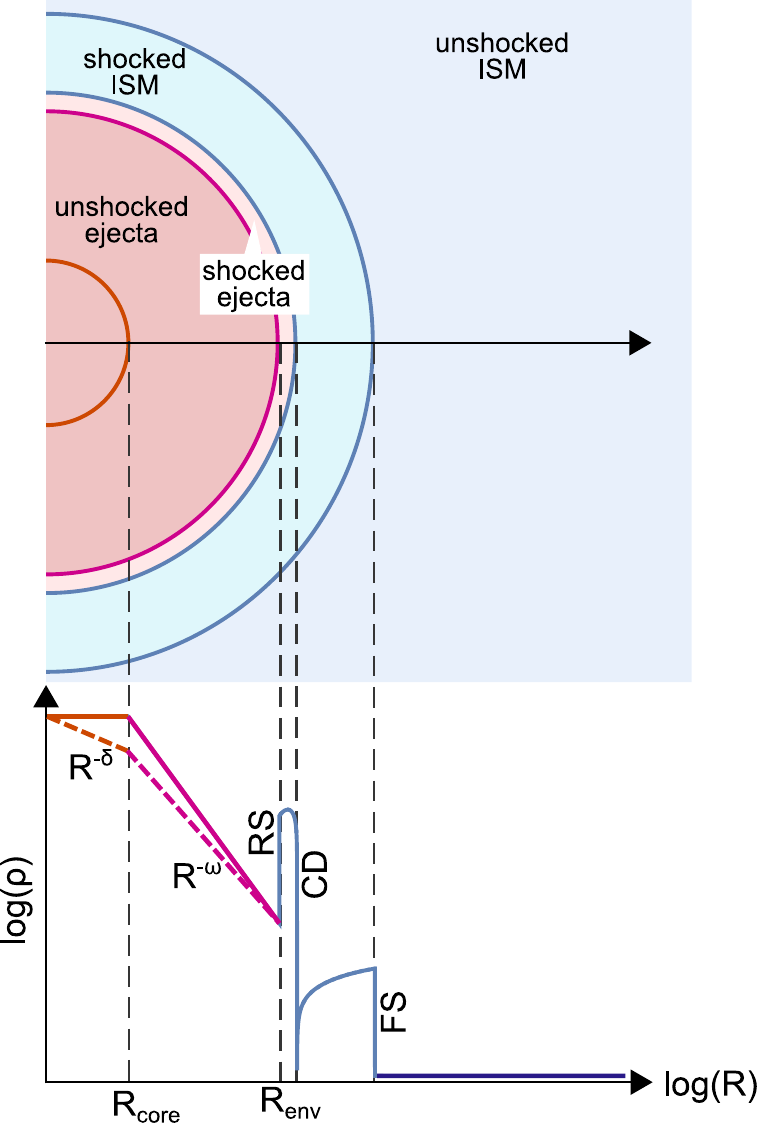}
    \caption{Cartoon of the initial SNR structure with focus on the radial profiles of the density in the core ($R<R_{\rm{core}}$) and envelope ($R_{\rm{core}}<R<R_{\rm{env}}$) of the ejecta.}
    \label{fig:SNRcartoon}	
\end{figure}
%
The cartoon of Fig.~\ref{fig:SNRcartoon} shows a schematic representation of the SNR structure, featuring the RS, CD and FS and the density in a qualitative way.
Typical models for the density profile of the freely expanding ejecta, where velocity scales linearly with radius, assume a core with a shallow radial profile $\propto r^{-\dl}$ (with $\dl<3$), plus an envelope with a steep power-law one $\propto r^{-\om}$ (with $\om>5$), according to:
\begin{equation}
\rhoej(r,t)=
\begin{cases}
A\,(\vt/r)^\dl/t^{3-\dl}, & \text{if } r < \vt t\,,	\\
A\,(\vt/r)^\om t^{\om-3}, & \text{if } \vt t \leq r < \Rrev\,,
\end{cases}
\label{eq:rhoejprofile}
\end{equation}
where the parameters $A$ and $\vt$ (the expansion velocity of the core boundary, in the unshocked ejecta) are given by:
\begin{eqnarray}
\label{eq:Adef}
\vt &=& \sqrt{\frac{2(5-\dl)(\om-5)}{(3-\dl)(\om-3)}\frac{\Esn}{\Mej}}\,,\\
A &=&\frac{(5-\dl)(\om-5)}{2 \pi (\om-\dl)}\frac{\Esn}{\vt^5}\,
\,.
\end{eqnarray}

Typical values for $\dl$ are found to vary in the range $0.001\lesssim \dl \lesssim 1$ for different types of supernovae, with indications that $\dl<1$ are characteristic of Type II SNe, while the outer envelope index is considered to vary in the range $7\lesssim \om \lesssim 12$ (\citealt{Chevalier:1989}, TM99, \citealt{Matzner:1999,Chevalier2005,Ferrand:2010,Bucciantini_Arons+11a,Kasen:2010,Miceli:2013,Potter2014,Karamehmetoglu:2017,Kurfurst:2020,Meyer:2020,Meyer2021}).

Note that models having the same $\om$ and $\dl$, once scaled with their respective characteristic scales ($\Rch, \tch$), form a family of equivalent solutions.


\section{Early evolution}
\label{sec:selfsim}
Before discussing the results of our numerical models, let us review and implement what is known about the very early phases.

C82 derived  self-similar solutions for the radial profiles of density, velocity, and pressure, in the initial phase of evolution, namely as long as the RS propagates through the envelope of the ejecta.
To avoid confusion, in the following we will specify where our notation differs from that of the reference paper. For simplicity in Appendix \ref{app:notations}, we recall the used notation and, whenever possible, we directly compare it with the one from TM99 and C82.

To summarize the approach in C82 for a homogeneous ambient medium, corresponding to the case $s=0$ in the notation of that paper, let us first note that, during these early phases, the evolution in the whole region between the RS and the FS must only depend on the dimensional quantities $\rhoism$, i.e. the density of the ambient medium, and $\rhoej r^{\om}t^{3-\om}$,  a (time- and space-independent) quantity that is related to the density of the expanding envelope,  is labelled in Sec.~\ref{sec:basic} as $A \vt^\om$, and corresponds to $g^n$ in the C82 notation; also note that in that paper our index $\om$ is called $n$. 
For instance, the total SN energy and mass of the ejecta do not enter here, because all the material inside the RS is causally disconnected.

Therefore one cannot derive independently the characteristic length and time for this problem, but they must be linked together.
First, the RS, CD, and FS sizes must all evolve like $t^{(\om-3)/\om}$.
In addition, apart from their respective dimensional scalings, the radial profiles of the hydrodynamic quantities must all depend on just one self-similar variable, comoving with the RS, CD, and FS.
Following C82, we choose here the self-similar independent variable: 
\begin{equation}
\eta\equiv r^\lmb/t, \hspace{.5cm} {\rm where} \hspace{.5cm} \lmb=\om/(\om-3).
\end{equation}
We can then factorize the hydrodynamic quantities (here velocity, sound speed and pressure) in a dimensional part, and in a dimensionless, self-similar one:
\begin{equation}
  v(r,t)=U(\eta)\frac{r}{t}\,;\quad
  c_s(r,t)=C(\eta)\frac{r}{t}\,;\quad
  p(r,t)=P(\eta)A\vt^\om\frac{r^{2-\om}}{t^{5-\om}}\,.
\end{equation}
The local sound speed can be written as $c_s=\sqrt{\Gamma p/\rho}$, where $\Gamma$ is the adiabatic index.
Differently from C82, we have preferred to use here a dimensionless quantity also for the pressure, by extracting explicitly $Av_t^\om$. Namely the definition of $P(\eta)$ is not the same as in C82.
Anyway, this different choice does not affect the final results.
The hydrodynamic equations can then be transformed in the following ordinary differential equations in the $\eta$ variable  (C82):
\begin{eqnarray}
U^2-U+(\lmb U-1)\eta U'+\left(\frac{\lmb}{\Gamma}\frac{\eta P'}{P} -\frac{\om-2}{\Gamma}\right)C^2&=&0\,;   \\
(\om-3)(1-U)+\lmb\eta U'+(\lmb U-1)\left(\frac{\eta P'}{P}-2\frac{\eta C'}{C}\right)&=&0\,;  \\
(\om-5)-\Gamma(\om-3)+\left(2+(\Gamma-1)\om\right)U
\quad\qquad\qquad&&\qquad
\nonumber\\
+(\lmb U-1)\left(2\Gamma\frac{\eta C'}{C}-(\Gamma-1)\frac{\eta P'}{P}\right)&=&0\,.
\end{eqnarray}
%
One may note that $\eta$ is explicitly present only together with a first derivative of some quantity (labelled with a prime), which means that the solutions are invariant by translation in $\eta$.
In addition, the self-similar pressure $P$ appears only in the form of a  logarithmic derivative, which means that its solution is invariant \RB{by} an arbitrary scaling.
For these reasons, the numerical solutions for the inner and outer sides to the CD can be integrated independently, starting from the RS and FS boundaries.
In a second stage one can fix the regularity conditions, by assigning the same $\eta$ value (conventionally 1) on both sides of the CD, as well as by imposing pressure continuity across the CD itself.
For completeness, the boundary conditions for the self-similar quantities read:
\begin{equation}
    U\rs{RS}=\frac{\Gamma-1+2/\lmb}{\Gamma+1}\,;\quad
    C^2\rs{RS}=\frac{2\Gamma(\Gamma-1)}{(\Gamma+1)^2}\left(1-\frac{1}{\lmb}\right)^2,
\end{equation}
in the downstream of the RS, while:
\begin{equation}
    U\rs{FS}=\frac{2}{\lmb(\Gamma+1)}\,;\quad
    C^2\rs{FS}=\frac{2\Gamma(\Gamma-1)}{(\Gamma+1)^2}\frac{\Gamma}{\lmb^2}\,,
\end{equation}
in the downstream of the FS.
Using the approach described above, we have computed the radial structure of the RS, CD and FS for a large number of cases, with values of $\om$ ranging from 6 to 100. 
Although the solutions for the positions of RS, CD, and FS are not analytical, we provide here highly-accurate analytic approximations for some of these quantities.
In comparison, C82 tabulated them only for a limited selection of cases.
The (exact) formula for the evolution of the CD radius ($\Rcont$) is:
\begin{equation}
    \Rcont(t)= \left(\frac{9\,A\,\vt^\om}{\al(\om)(\om-3)^2\rhoism}\right)^{1/\om} t^{(\om-3)/\om}\,,
    \label{eq:RcontCh}
\end{equation}
with:
\begin{equation}
\al(\om)\equiv\frac{p\rs{RS}}{p\rs{FS}}\left(\frac{\Rforw}{\Rrev}\right)^2\left(\frac{\Rrev}{\Rcont}\right)^\om,
\end{equation}
where $p_{\rm{RS}}$ and $p_{\rm{FS}}$ are the pressures at the two shocks.
Unfortunately there is no analytic solution for this quantity, but it can be accurately approximated by:
\begin{equation}
    \al(\om)\simeq\frac{0.79966-0.49408\sqrt{\om-5}+0.68648\,(\om-5)}{2.03247-0.63043\sqrt{\om-5}+(\om-5)}\,.
\end{equation}
Using this formula the quantity $\al^{-1/\om}$, which enters in Eq.~\ref{eq:RcontCh}, is approximated to better than 0.01\% for all $\om$ values larger than 6.
The positions of the RS and FS radii ($\Rrev$ and $\Rforw$ respectively) with respect to that of the CD can be approximated well by: 
\begin{equation}
   \frac{\Rrev}{\Rcont}\simeq1-\frac{0.21064\left(1+0.06245/\sqrt{\om-5}\right)}{1.38208+(\om-5)}\,,
\end{equation}
and:
\begin{equation}\label{eq:Rfs_Rcd}
   \frac{\Rforw}{\Rcont}\simeq1.09572+\frac{0.18326}{0.14675+(\om-5)}\,.
\end{equation}
It can be seen that, for very large $\om$ values, $\Rforw/\Rcont$ reaches an asymptotic value $\simeq1.09572$.
Both approximations reach an accuracy of 0.003\% for $\om>6$.
Note however that all the formulae given above are not reliable for $\om$ values approaching the critical case $\om=5$.
During this self-similar phase, the mass of the ejecta collected by the RS is:
\begin{eqnarray}
    M(t)&=&\int_{\Rrev(t)}^\infty4\upi A\vt^\om r^{-\om} t^{\om-3} r^2 dr \nonumber\\ 
    &=&\frac{4\upi A\,\vt^\om}{\om-3}\left(\frac{\Rrev(t)}{t}\right)^{(3-\om)}
    \!\!\!\!
    \propto (t^{(\om-3)/\om-1})^{3-\om}=t^{3(\om-3)/\om}\,,
\end{eqnarray}
and therefore the time at which the RS reaches the core of the ejecta, namely when $M(t)$ becomes equal to the total mass of the envelope  $M_{\rm{env}}=(3-\dl)\Mej/(\om-\dl)$, is found to be:
\begin{equation}
\label{eq:tcore}
    \frac{\tcore}{\tch}=\left(\frac{81 (3-\dl)^5(\om-3)(\om-5)^{-3}}{128\,\upi^2\alpha^2(\om)(5-\dl)^3(\om-\dl)^2}\right)^{1/6}
    \left(\frac{\Rrev}{\Rcont}\right)^{\om/3}.
\end{equation}
After this time, the RS shock no longer follows the power-law expansion characteristic of the self-similar solution: its further evolution will be more complex, requiring a fully numerical investigation.

\begin{figure}
\centering
	\includegraphics[width=.48\textwidth]{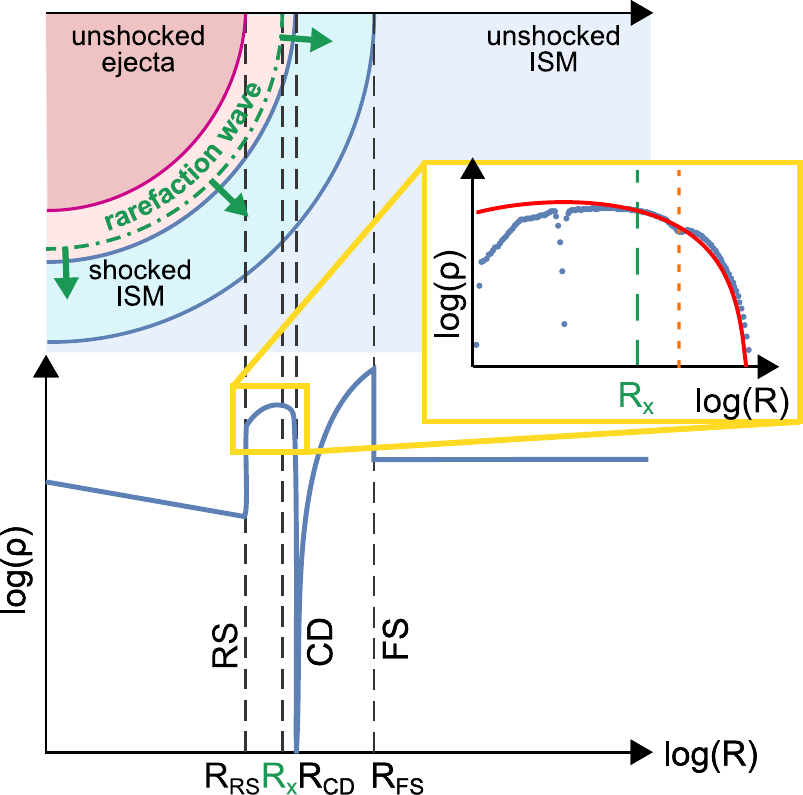}
    \caption{Cartoon illustrating the position $R_x$ at a particular time of the rarefaction wave (in green), propagating  the information of the RS interaction with the core in the shocked ejecta. Colors are the same as in Fig.~\ref{fig:SNRcartoon}. In the yellow box we show a zoom-in of the effect of the rarefaction wave on the ejecta structure: blue points are from the simulation, in red the comparison with the predicted structure from the pure Chevalier model (no rarefaction wave). The initial boundary of the core is indicated by the dotted orange line.
    It can be appreciated how the passing of the wave produces a lowering and widening of the ejecta profile. Notice that the real relative dimensions of the radial profile in the main figure are not maintained in order to better appreciate the global structure.}
    \label{fig:rareF}	
\end{figure}
The RS and CD will experience the end of the self-similar solution as well, but at later times.
This is because the information about the change of the ejecta density profile will take some time to reach the outer regions.
After $\tcore$, a rarefaction wave propagates at the sound speed through the shocked layers (see Fig.~\ref{fig:rareF}), so that the delayed time $t_x$ for the wave reaching a radial distance $R_x$ (corresponding to the self-similar coordinate $\eta_x$), can be evaluated as:
\begin{equation}
    \frac{t_x}{\tcore}=\exp\left(\int_{\eta\rs{RS}}^{\eta_x}\frac{d\eta}{\eta\left[\lmb(C(\eta)+U(\eta))-1\right]}\right),
\end{equation}
Let us call $\tcoreCD$ and $\tcoreFS$ the times at which the sound wave has reached respectively the CD and the FS.
For these quantities we have derived the following approximated functions:
\begin{eqnarray}
\frac{\tcoreCD}{\tcore}\!\!\!\!\!\!&\simeq&\!\!\!\!\!\!1.10672+\frac{0.37713}{1.50122+(\om-5)}, \label{eq:tcoreCD}\\
\frac{\tcoreFS}{\tcore}\!\!\!\!\!\!&\simeq&\!\!\!\!\!\!1.35730+\frac{1.67250}{0.27501(\om-5)^{0.13135}+(\om-5)}, \label{eq:tcoreFS}
\end{eqnarray}
both with an accuracy better than 0.01\%, for $\om>6$.
One should notice that C82 solutions are by themselves independent of the density profile in the core, since it has not been reached as of yet.
The presence of $\dl$ in Eq.~\ref{eq:tcore} comes only from the fact that the fraction of mass in the envelope depends also on the density profile in the core.
For $\om$ approaching infinity, $\tcore$ vanishes as $\om^{-2/3}$.

In the following we shall also consider the limit case of $\om=\infty$, which corresponds to ejecta without any envelope, and is dubbed by TM99 as $\om=0$.
Following the above equations, in that limit we found $\tcore$ to vanish, as well as the distance between the RS and the CD.
%
The last result can be easily understood, having in mind that all the ambient medium swept by the FS will then be stored in the volume between the CD and the FS itself, so that:
\begin{equation}
    \frac{4\upi}{3}\rho_0\Rforw^3= \frac{4\upi}{3}\rho\rs{avg}\left(\Rforw^3-\Rcont^3\right)\,
\end{equation}
where $\rho\rs{avg}$ is the average density in the shell bound by the CD and FS.
In the asymptotic limit $\om \rightarrow \infty$, from the value of the ratio $\Rforw/\Rcont\simeq 1.09572$ (Eq.~\ref{eq:Rfs_Rcd}) it can be derived $\rho\rs{avg}=4.16933 \,\rho_0$, actually very close to the factor 4 right behind the FS.

\section{Numerical Scheme}
\label{sec:numerical}
In this section we present the hydrodynamic Lagrangian code that we have developed to model numerically the one-dimensional hydrodynamic evolution of SNRs.
One advantage of using a Lagrangian scheme over an Eulerian one is that one can follow in greater detail the structure of the SNR envelope even in those regions with steep gradients, like in the envelope of the SNR ejecta, preserving with high accuracy the large density discontinuities.
In addition, the ability to follow the fluid elements as they expand, contrary to Eulerian scheme on fixed grid, relaxes the computational requirement of a large dynamical range in radius.

We have developed a 1D Lagrangian hydrodynamical scheme, following the recipes described in \citet{Mezzacappa_Bruenn+93a}.
Shocks are handled through the implementation of standard von Neumann-Richtmyer viscosity \citep{Von-Neumann_Richtmyer50a}.
For the reader convenience, we briefly summarize here the equations that are solved.
In order to gain accuracy in time, without resorting to a full second-order approach, the time evolution of the velocity ($v$) and radius($r$) of interfaces {\footnotesize{$i+1/2$}} between the  shell {\footnotesize{$i$}} and the shell {\footnotesize{$i+1$}} has been modified, with respect to the standard  half-step time staggering, to account also for their acceleration ($a$), according to:
\begin{align}
 Q_{i}^n    \quad           =& \; \eta_{\rm vnr} \rho_i^n( v_{i+1/2}^{n}-  v_{i-1/2}^{n})^2\Theta[v_{i-1/2}^{n} -v_{i+1/2}^{n}]\,,\\
  a_{i+1/2}^{n}     =& \;  [(r_{i+1/2}^n)^2(p^n_{i+1}-p^n_{i}) - (r_{i+1}^n)^2Q_{i+1}^n + (r_{i}^n)^2Q_{i}^n]/\Delta m_{i+1/2}\,,\! \\
  v_{i+1/2}^{n+1} = & \;  v_{i+1/2}^{n} - 4\pi \Delta t   a_{i+1/2}^{n}\,, \\
  r_{i+1/2}^{n+1}   = & \; r_{i+1/2}^{n} + 4\pi \Delta t v_{i+1/2}^{n} + 2\pi (\Delta t )^2 a_{i+1/2}^{n} \,,
\end{align}    
where $\Theta [\cdot] $ is the Heavyside function, $\Delta t$ is the time interval between the steps $n$ and $n+1$ (chosen in order to ensure a stable time evolution based on the standard Courant–Friedrichs–Lewy condition, and the requirement that no shell should compress or expand  more than a factor 1.2 in each time step),
$\Delta m_{i+1/2}$ is the mass at the interface, defined as a function of the mass of the two bounding shells $\Delta m_{i+1/2}=(\Delta m_{i+1}+\Delta m_{i})/2$, $\eta_{\rm vnr} = 4$ is the viscosity coefficient (which we chose larger than the typical value $2$ of the standard von Neumann-Richtmyer method in order to suppress numerical noise arising from the presence of strong shocks) and $Q$ is the viscous pressure \citep{Schulz:1964}.
The radius of each shell is defined as the barycenter radius:
\begin{align}  
r^n_i = \left(\frac{ (r_{i+1/2}^n)^3+  (r_{i-1/2}^n)^3}{2}\right)^{1/3}\,,
\end{align}  
and its density:
\begin{align}  
\rho^n_i = \frac{3 \Delta m_{i}}{4\pi(  (r_{i+1/2}^n)^3-  (r_{i-1/2}^n)^3)}\,.
\end{align}  
Instead, the pressure is derived by solving (either by successive iterations or by direct solution) the following equation for the specific internal energy $e_i$:
\begin{align}
e^{n+1}_i &= e^{n}_i -\frac{p^{n+1}_i  +p^{n}_i}{2} \left(\frac{1}{\rho^{n+1}_i }-\frac{1}{\rho^n_i } \right)\nonumber+\\
		& \quad-2\pi \Delta t [r^{n+1}_i  +r^{n}_i]^2 Q_i^n\frac{v_{i+1/2}^{n}-v_{i-1/2}^{n}}{\Delta m_{i}}\,,
\end{align}
and assuming the following equation of state $p_i =2 \rho_i e_i/3$, appropriate for a perfect gas of adiabatic index $\Gamma = 5/3$.
In order to avoid spurious entropy generation associated to numerical noise, we force the entropy to remain constant in the region bounded by the RS and FS.
Our numerical models are initialized using the analytical SNR solutions of C82 hereafter (see Section~\ref{sec:selfsim}), at a time corresponding to $0.9 \tcore$.
Our resolution is set in order to have 500 mass shells uniformly spaced in radius between the initial positions of the RS and FS.
The unshocked cold ejecta inside the RS are resolved over 4000 mass shells logarithmically-spaced in radius up to the center.
The unshocked cold ISM outside the FS is resolved over 1500 logarithmically-spaced shells up to an outer radius of $11 \Rch$.
The stretching in these two regions is chosen in order to have a smooth change in resolution at the RS and FS locations.
We have verified by either changing the initial time to 
$0.5 \tcore$, and by doubling the resolution in each zone, that results are unchanged and that the numerical model preserves the C82 structure up to $\tcore$.

The boundary conditions at the outer radius pose no problem if one chooses it far enough so that for the SNR FS never reaches it during the evolution.
As inner boundary condition the first interface is held fixed at the center $r_{1/2}=0$, and reflection is imposed on the other fields.
This condition is robust and allows the late bouncing of the RS without artifacts.

In order to treat the asymptotic case $\om=\infty$, we have used a different initialization for our simulations, maintaining the same numerical scheme and grid, but based on the \citet{Parker:1963} model.
This is equivalent to the C82 one for the shocked ambient medium, except for the fact that now one considers a spherical piston, instead of the CD, moving like $t^{1/\lmb}$, where the link to the C82 problem is obtained by setting $\lmb=\om/(\om-3)$. 
In this case the initial time of the simulations is set to $\tini=0.01\tch$.
%

\section{Analysis of the numerical results}
\label{sec:snr}

We have used our Lagrangian code (described in the previous section) to analyze multiple possible properties of the ejecta, shaping differently their core and envelope, characterized respectively by the parameters $\dl$ and $\om$.
We have explicitly calculated models with $\om=6$, 7, 8, 9, 10, 11, 12, 14, 18, 25, 50, $\infty$ (the choice of these values has been simply motivated by the need to provide a suitably spaced sampling for our interpolations and for a direct comparison with TM99 results).
As already mentioned before, in each case with a finite $\om$ value, we have taken as a starting time of our numerical models the value $\tini=0.9\,\tcore$.
For the shocked region we have used the Chevalier's solution (C82).
Only in the $\om=\infty$ case we have used $\tini=0.01\tch$, and an initial profile without the shocked ejecta and with a shocked ambient medium, according to a Parker's solution.
We have repeated all the simulations for different values of the core power law indices in the range of interest, namely $\dl=0$, 0.1, 0.5, 1.0.

In order to test the numerical convergence of our models, we have also run simulations in which the density profile of the ejecta has been cut at a radius at which the density was comparable with that of the ambient medium, and the ejecta have been left at the initial time in direct contact to the ISM, without any shocked medium in between.
We have run different cases with varying $t_{\rm{ini}}$, and we found a reasonable convergence of the solution for $t_{\rm{ini}}\leq 0.01\tch$.
Some noticeable difference in their late evolution appeared for small $\om$ values, simply because the arbitrary cut in the profile of the ejecta implies a modification of both total mass and energy.
We point out, however, that these simulations have been run only to test the stability of the numerical code; while for the results that follow we have used those initialized with the self-similar solutions.

\begin{figure}
\centering
	\includegraphics[width=.47\textwidth]{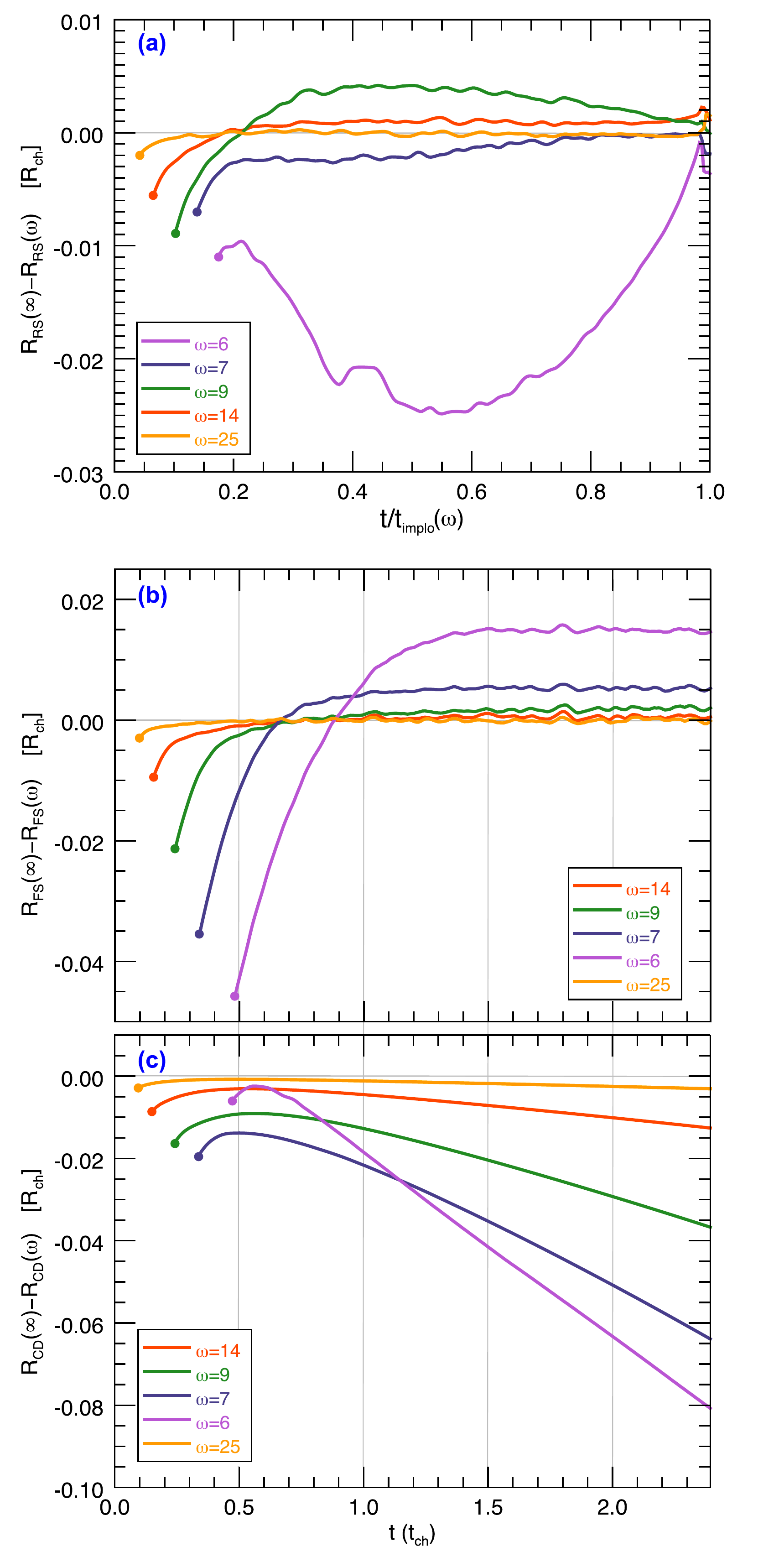}
    \caption{Convergence of the RS (a), FS (b) and CD (c) curves to the asymptotic case $\om=\infty$. Depicted are cases with $\om=6,\, 7,\, 9,\, 14,\, 25$ and $\dl=0$.
    The convergence is shown as differences between each of the curves with respect to the asymptotic one and as function of time normalized to the  implosion time $\timplo(\om)$ in the RS case, or as function of time in characteristic units. 
    Each curve is plotted starting from $t=\tcore$.
    Notice that the percentage difference between the curves is really small, and the convergence is indeed fast with increasing the $\om$ value.
    It should not be surprising that the behaviour of $\om=6$ diverges so much from the other ones, since it is so close to the critical case $\om=5$. And a similar effect can be noticed also in the following figures.
    }
    \label{fig:traj_conv}	
\end{figure}

\begin{figure*}
\centering
	\includegraphics[width=.99\textwidth]{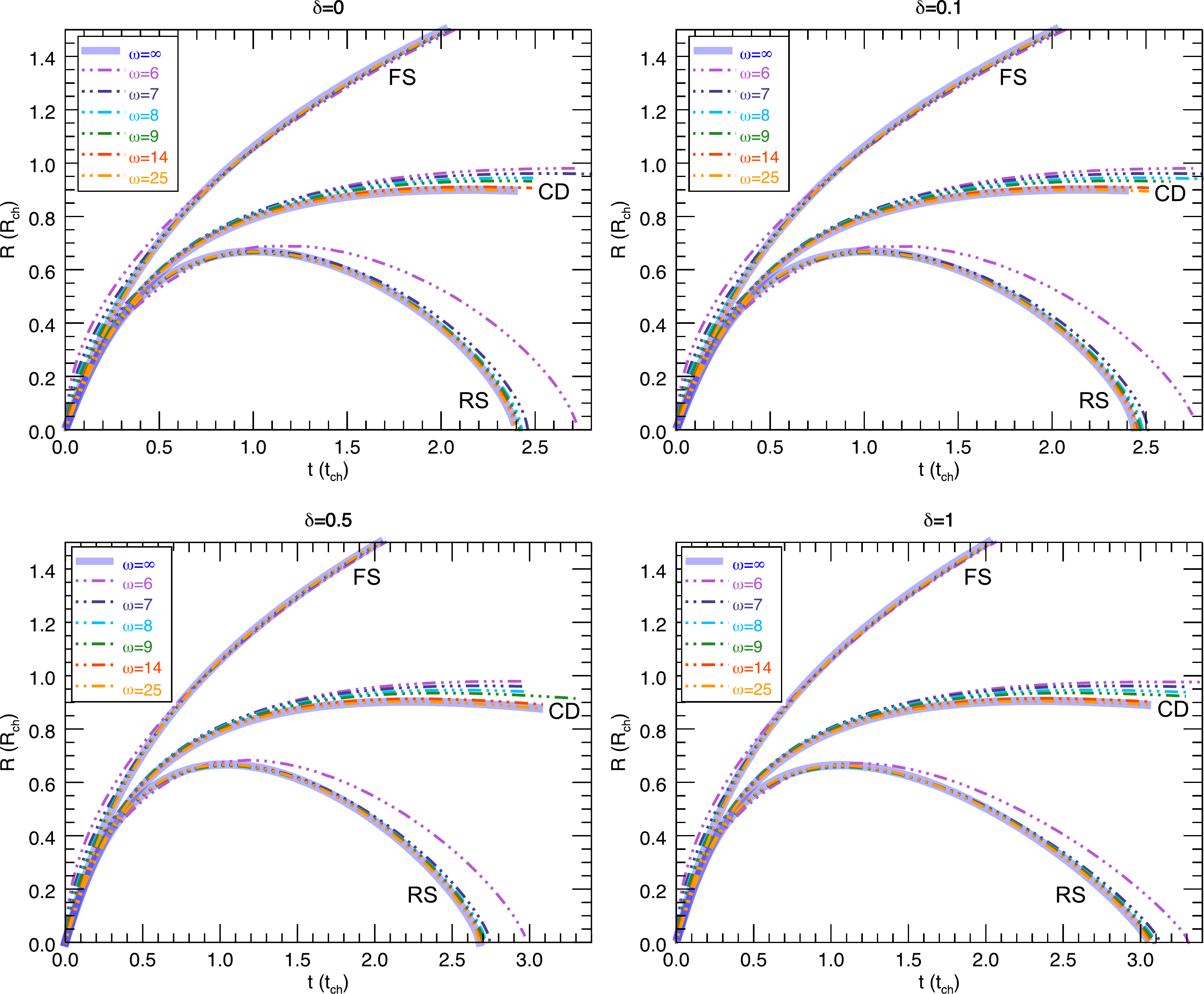}\,
    \caption{Evolutionary paths of the FS, CD and RS as extracted from the results of our numerical simulations for the cases $\om=6,\,7,\,8,\,9,\,14,\,50,\,\infty$, and for different values of the core index $\dl$, namely: $\dl=0$ (top-left), $\dl=0.1$ (top-right), $\dl=0.5$ (bottom-left) and $\dl=1 $ (bottom-right). Notice that the x-axis scale is different in the upper and bottom panels. For $t\leq\tcore$ we plot the C82 analytic profiles presented in Sec.~\ref{sec:selfsim}.}
    \label{fig:numerical_traj_allOM_D}	
\end{figure*}

Fig.~\ref{fig:traj_conv} shows our first result: the convergence of the evolutionary curves of the RS, FS and CD to the asymptotic ones ($\om=\infty$), for a representative selection of $\om$ values ($\dl$ is set here to 0).
The convergence of the RS (panel (a)) is shown as function of the time scaled with the time at which the RS implodes ($\timplo(\om)$).
%
%
%
When increasing the $\om$ value, the curves approach the asymptotic one with a monotonic behavior.
We notice that the convergence of the RS trajectory to the asymptotic one
with the $\om$ value is rather fast.
A deviation of less than $0.01\,\Rch$ at $t=0.5\,\timplo$ can be found starting from $\om=14$, while for $\om=25$ the deviation is non-significant and the two curves appear to be coincident.
The same conclusion remains valid also for the $\dl>0$ cases, for which convergence at large $\om$ values is even more rapid than in the flat case (as it can be appreciated if comparing the different panels of Fig.~\ref{fig:numerical_traj_allOM_D}).

Fig.~\ref{fig:traj_conv} also shows the convergence to the asymptotic case of the CD and FS trajectories, for the same selection of $\om$ values and in the $\dl=0$ case.
We observe the FS to converge very quickly to its asymptotic limit when increasing $\om$, with the two curves becoming almost coincident from $\om=14$, while in the $\om=9$ case the difference is only of a few $10^{-3}$ up to the implosion time.
We notice that the differences in the FS curves are  mostly limited to the initial part of the evolution (say for $t\lesssim 1.5\,\tch$), while at later times they all show the same trend and are separated by the asymptotic curve only by a constant shift (the maximum is of $0.015\,\Rch$ for $\om=6$).
Finally the CD appears to be the slower curve to converge to the asymptotic profile with the $\om$ value, with a small difference of $\sim 1.5\times10^{-3}$  still noticeable for  $\om=25$ at $t=2\,\tch$.

Comparing the convergence of all the curves shown in Fig.~\ref{fig:traj_conv}, it is quite clear that the asymptotic limit is rapidly reached while increasing $\om$, with the differences between $\om=25$ and $\om=50$ being only of a few $\times 10^{-3}$, and becoming even smaller for higher $\om$ values.

As advanced above, the most sensitive curve to the variation of the model parameters, except for the RS in the $\om=6$ case, is found to be the CD, as it can also be appreciated looking at Fig.~\ref{fig:numerical_traj_allOM_D}. 
Here we show a direct comparison of the FS, CD and RS trajectories with varying $\om$ (same panel) and for the different values of $\dl$ considered (different panels).
Namely we plot the curves for the cases: $\om=6$, 7, 8, 9, 14, $\infty$  and $\dl=0$, 0.1, 0.5, and 1.0.
We can notice again the monotonic convergence of the curves to the asymptotic one, which is very fast  for the FS.
The FS and CD appear to be very weakly affected by the core structure, thus their main properties can be described only based on the $\om$ family.
On the contrary, the RS shows a relevant deviation from the flat case ($\dl=0$), already for $\dl>0.1$, with the structure of the core reflecting in a slower propagation of the shock towards the center, and a modification of the implosion time from $[2.39-2.71]\,\tch$ for the $\dl=0$ case (considering all the $\om$ values) to the $[3.1-3.3]\,\tch$ for the $\dl=1$ one.
It can be also seen that, augmenting $\dl$, the differences between the RS for the lower and higher $\om$ values tend to diminish, with a faster convergence to the asymptotic case at higher $\dl$.
On the other hand, the difference in the implosion time between the flat case and $\dl=0.1$ is only of $5\%$ for all the $\om$ values, while for $\dl<0.1$ this difference becomes irrelevant.
This is the reason why we will consider the case $\dl=0$ for comparing directly with TM99 results.
They in fact assume $\dl=0$ in general, while a slightly higher value of $\dl=0.03$ is used when the flat limit is considered not to be a good approximation. 
Following our results, this latter case can be safely assimilated to the flat one, without introducing any significant deviation to the SNR characteristic curves.

\begin{figure*}
\centering
	\includegraphics[width=.99\textwidth]{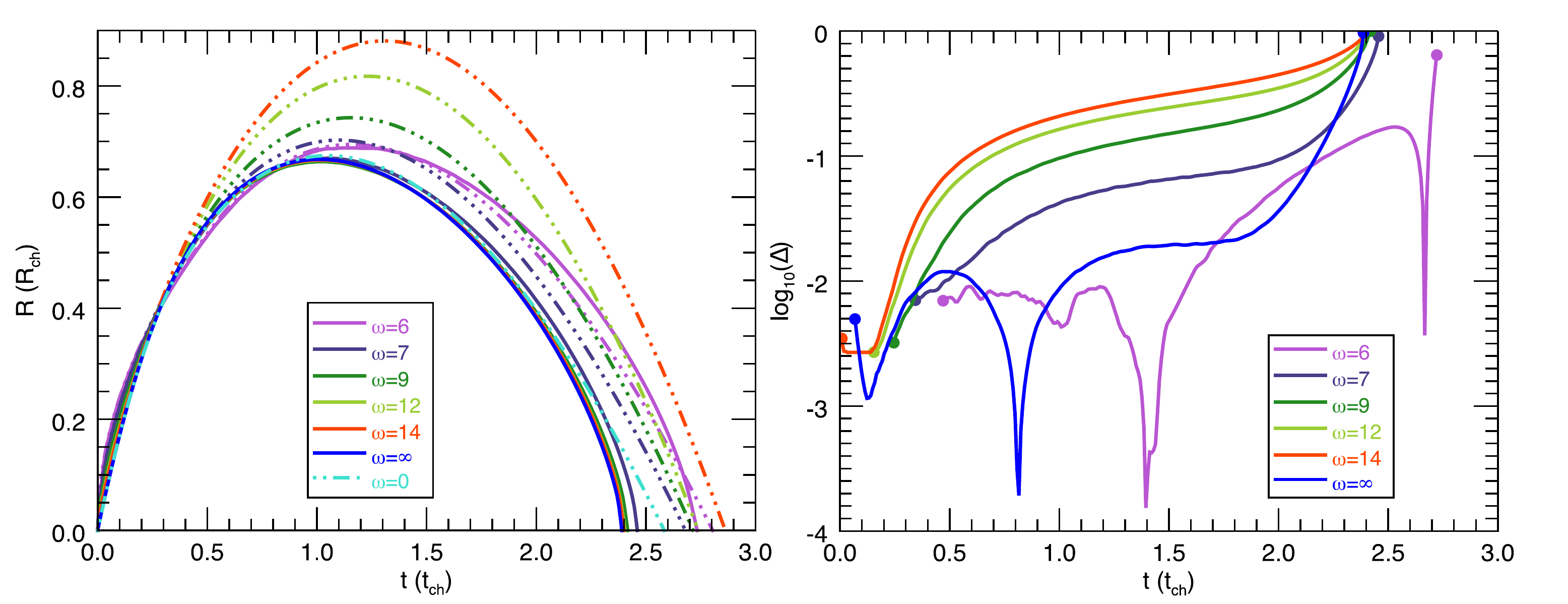}
    \caption{\textit{Left panel}: Direct comparison of the RS trajectories for different $\om$ values as computed with our numerical models (solid lines) and from the semi-analytic prescriptions of TM99 (dot-dashed lines). The numerical results are given for the $\dl=0$ case, as is the most suitable for a comparison with TM99 assumptions.
    \textit{Right panel}: 
    Plot of the logarithm of the relative difference between the curves in the TM99 approximation and this work, namely: $\Delta=\abs{R_{\mathrm{RS}}^{\mathrm{TM99}}(\om) - R_{\mathrm{RS}}(\om) }/ R_{\mathrm{RS}}^{\mathrm{TM99}}(\om) $. The curves are only plotted in the range $\tcore \leq t \leq \timplo$ (from the numerical models). 
    The large dips in these curves happens because of the crossing of the solutions (where the log relative difference is formally $-\infty$).}
    \label{fig:cfr_NUM_true}
\end{figure*}

Fig.~\ref{fig:cfr_NUM_true} shows a direct comparison of the trajectories of the RS as obtained with our numerical simulations with those computed using TM99's formulas.

We show a selection of $\om$ values, namely: $\om=6$, 7, 9, 12, $\infty$ for the $\dl=0$ case that, as already pointed out, is the most suitable for this comparison.
Let us first focus on the asymptotic curves ($\om=\infty$ in our notation, equivalent to the case labelled as $\om=0$ in TM99 one).
The two differ only by $\sim 1\%$ around the RS maximum ($t\sim\tch$).
On the other hand, the approximations that TM99 introduced to get a semi-analytic description of the RS trajectory cause an important deviation close to the implosion time, with an evident difference in the slope of the RS for $t\gtrsim 0.8\,\timplo$ ($\gtrsim2.1\,\tch$ in this case). 
The same can be actually observed if comparing the original numerical and semi-analytical results of TM99.
We conclude that this deviation is a consequence of the chosen approximations and not of the numerical model itself.
The only other case in which we found some  coincidence of the TM99 solutions with our results is that of $\om=6$.
But when increasing the $\om$ value, the difference between the trajectories clearly increases.
For instance, in TM99 the maximum size of the RS, for $\om$ changing from 9 to 12, changes from $0.743\,\Rch$ to $0.881\,\Rch$, while in our model the variation is very small, from $0.664\,\Rch$ to $0.666\,\Rch$.
We found that the relative variation between the models, $\Delta=\abs{R_{\mathrm{RS}}^{\mathrm{TM99}}(\om) - R_{\mathrm{RS}}(\om) }/ R_{\mathrm{RS}}^{\mathrm{TM99}}(\om)$, is typically of the order of 10\%. It can easily go much higher, as for instance, it is 30\% in the $\om=14$ case at time $t=1.5\,\tch$, as can be seen in Fig.~\ref{fig:cfr_NUM_true}. 
Also, as already noticed, the loss of accuracy of the chosen representation of the curves in the TM99 approximation leads to even larger relative differences closer to the implosion time, which are close to 100\%.
Moreover it is clear that the TM99 solutions do not converge, for large $\om$ values, to their asymptotic solution $\om=0$ (equivalent to $\om=\infty$ in our notation), for which they get $0.674\,\Rch$.
For comparison, the value we get for the asymptotic case is instead $0.668\,\Rch$.

\begin{figure}
\centering
	\includegraphics[width=.5\textwidth]{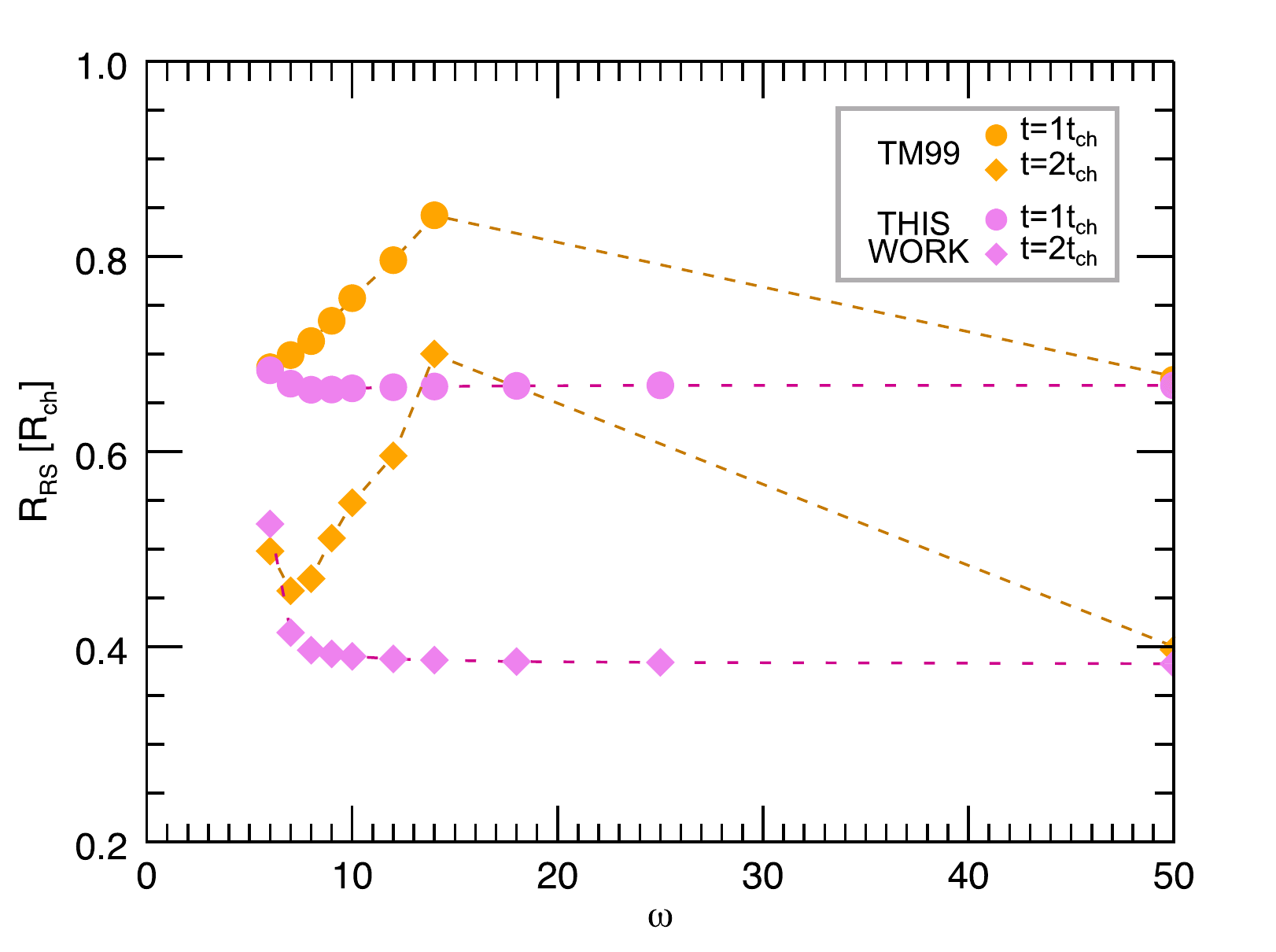}
    \caption{Values of the RS radius at fixed time ($t=1.0\,\tch$ with circles and $t=2.0\,\tch$ with diamonds) as function of the $\om$ value, varying in the range $\om=6$, 7, 8, 9, 10, 12, 14. 
    Our results are shown in pink color, while those from TM99 in orange color.
    As reference for the asymptotic case we have used the model with $\om=50$, in correspondence to which we have plotted the asymptotic case as from TM99.
    The dashed lines must not be considered as fits of the data, but simply represent a graphical connection of points to help visualizing their trend.
    }
    \label{fig:cfr_NUM_true2}
\end{figure}

 This can be more easily appreciated in Fig.~\ref{fig:cfr_NUM_true2}, where we show the radius of the RS for a selection of $\om$ values and at fixed times ($t=1.0\,\tch$ and $t=2.0\,\tch$), in comparison with the results obtained using the TM99 formulae. Again, it is evident that the TM99 solutions are not converging to the asymptotic value $\om=\infty$. On the contrary, they appear to diverge while increasing $\om$, meaning that the maximum radius of the RS becomes larger and larger when increasing the steepness of the density profile of the ejecta envelope. This does not appear to be physically motivated, since cases with a very high-$\om$ should approach that with an infinitely steep envelope. Our solutions are instead found to converge rather fast to the asymptotic case with increasing $\om$, what is true even if looking at different points of the evolutionary curve of the RS.

%

To summarize, our solutions show a rather fast monotonic convergence of the RS trajectories to the $\om=\infty$ one, so that all cases with $\om>9$ do not differ by more than 1\% along most of the trajectory.
For this reason the asymptotic case $\om=\infty$ can be used as an excellent approximation to model SNRs with steep density profiles. 
On the contrary we found that the accuracy of the widely used TM99 approximated expressions for the shocks evolution to be rather poor in some cases, especially for the RS, when changing the profile of the ejecta $\om$.
%

\section{Formulae for the FS-CD-RS trajectories}
\label{sec:traj}
To our understanding, there is no analytic nor semi-analytic treatment able to provide accurate formulae for the position of the shocks, as well as for that of the CD, when the conditions of validity
of C82's solutions no longer hold.
For this reason, while achieving the following results, we do not try to provide a detailed physical justification of the functional structure of our formulae, but simply to fit well the numerical results in a smooth way, with some further constraints that allow us to satisfy the asymptotic limits.
We will also discuss some details of the numerical results, trying to outline the underlying physics, even though in most cases just in a qualitative way.

As in the rest of this paper, we use here the dimensional scaling introduced in Sec.~\ref{sec:basic} (namely radii and times are expressed in terms of the characteristic quantities $\Rch$ and $\tch$).
Our formulae will then form a family, depending on the two ejecta parameters $(\om,\dl)$.
On the other hand, as we have seen in Sec.~\ref{sec:selfsim}, the C82 solutions, to which ours must connect, depend essentially just on the parameter $\om$, while the value of $\dl$ enters only  in determining the scaling for the fraction of mass in the envelope, and then in deriving from it the value of $\tcore$.
While our formula for the RS must match the C82 solution reasonably well at $\tcore$, our formulae for the CD and FS will have to match the corresponding C82 solutions at delayed times, respectively $\tcoreCD$ and $\tcoreFS$, which have been discussed and evaluated in Sec.~\ref{sec:selfsim}.

Our models have the following validity in terms of time: the RS trajectory is described with high accuracy (see the following subsection) up to the implosion time ($\timplo$), after which the RS disappears. 
The fit presented here for the CD trajectory is only valid up to $\timplo$: as we will discuss in the appropriate section, at $t>\timplo$ the CD starts to oscillate before reaching an asymptotic trend at very late times.
Finally the fit for the FS trajectory is valid up to late times since it is defined taking into account the asymptotic trend.

In the following subsections we will provide highly-accurate fitting formulae for the RS, CD and FS. 
In Appendix \ref{app:genTAB+cfrTM99} we provide some useful tables, giving respectively the summary of our general formulas (Table~\ref{tab:tabGEN}), and a direct comparison with the TM99 expressions (only valid for $\dl=0$; Table~\ref{tab:TM99vsUS}).

\subsection{Reverse Shock}
%
We have already discussed how the curves describing the evolution of the RS change their shape with $\om$ in a rather continuous and monotonic way (see Fig.~\ref{fig:numerical_traj_allOM_D}).
We have found that the similarity of the curves, at fixed $\dl$, can be better appreciated when scaling the time with $\timplo$: some noticeable difference can be observed only at low $\om$ values, while the curves are almost coincident for $\om\ge 8$.
At a given $\dl$ value, the main difference between curves with different $\om$ can be assimilated to a mere difference in the radial scaling.
This behaviour, at least, holds in the latter part of the RS evolution, when its size is contracting.
It can be understood with arguments similar to those used by \citet{Guderley:1942} to derive the  self-similar solutions for a converging shock: one may expect that in the late phases, just before $\timplo$, the RS implosion proceeds following a power-law evolution $\propto(t-\timplo)^\bt$, where $\bt$ is a function of only the central density profile (described by the parameter $\dl$).
From our numerical data we have extracted the following expression:
\begin{equation}\label{eq:Bguderlay}
\btimplo(\dl)=0.6824+0.07720\,\dl+0.02456\,\dl^2\,,
\end{equation}
that fits the data, in the range $0\leq\dl\leq1$. 
We have verified that $\btimplo(0)=0.6824$ is a very good approximation (with 1\% precision) of the exact value derived from the self-similar analysis (namely 0.68838).

On the other side, all the details of the former RS evolution concur in determining the time of implosion and the strength of the imploding RS.
As for the value of $\timplo(\om,\dl)$, we have derived the following analytic approximation:
\begin{equation}\label{eq:timplo}
\timplo(\om,\dl)=t^{\infty}_{\mathrm{implo}}(\dl) +\!\! \sqrt{\left(\frac{a_t(\dl)}{\om-5}\right)^2 \!\!\! + \left(\frac{-b_t(\dl)+c_t(\dl)/(\om-5)}{1+(\om-5)^2}\right)^2}\!,
\end{equation}
where the first term 
\begin{equation}
t^{\infty}_{\mathrm{implo}}(\dl)= a_{t}^{\infty} + b_{t}^{\infty}\,\dl + c_{t}^{\infty}\,\dl^2, 
\end{equation}
gives the approximated implosion time (for each $\dl$) in the asymptotic $\om$ limit, with an accuracy always better than 0.4\% for $\om\ge 14$, for all $\dl$. 
It can be noticed that $t^{\infty}_{\mathrm{implo}}$ increases with increasing $\dl$, 
a reasonable behavior having in mind that larger $\dl$ means higher densities near the core center, and therefore a more pronounced slowing down of the converging shock.
%
\begin{table}
\small
\centering
\caption{Coefficients of the implosion time ($\timplo$) general expression.}
\label{tab:PAR_timplo}
\begin{center}
\begin{tabular}{cc}
    \hline
    Function & Coefficient\\
    \hline
    \multirow{3}{*}{$t_{i}^{\infty}$} & $a_{t}^{\infty}= 2.399$\\
    &  $b_{t}= 0.4813$\\
    & $c_{t} = 0.1760$\\
    \hline
    \multirow{3}{*}{$\timplo$} & $a_t(\dl)=0.1006 + 0.04184 \,\dl$\\
    & $b_t(\dl) = 0.06494 + 0.09363 \,\dl$\\
    & $c_t(\dl) = 0.7063 -0.09444 \,\dl$ \\
    \hline
\end{tabular}
\end{center}
\end{table}
At a fixed $\dl$, $\timplo$ always increases with decreasing $\om$. The reason for this behaviour is that in not so steep density profiles of the envelope,  the RS must travel across more mass before reaching the core; an effect that is already apparent in the functional dependence of $\tcore$.
The values for the fit parameters are reported in Table~\ref{tab:PAR_timplo}.
The radial scaling for the RS, instead, shows a non-monotonic behaviour with $\om$: at large $\om$ values this scaling slightly decreases with decreasing $\om$; while at $\om\lesssim9$ the radial scaling increases again.

For the approximating formula we have then chosen to separate the time dependence from that on $\om$, namely:
\begin{equation}\label{eq:R_RS_FIT}
R_{\rm{RS}}(x,\delta,\omega)=\mathcal{R}(x,\delta)\times \mathcal{F}(\om, \delta)\,,
\end{equation}
where 
\begin{equation}
x=t/\timplo \hspace{0.5cm} {\rm and} \hspace{0.5cm}  \tcore < t \leq \timplo.
\end{equation}
The function $\mathcal{R}(x,\delta)$ represents a sort of universal shape of the RS trajectory in the asymptotic regime ($\om\rightarrow \infty$), valid for all the $\delta$ in the considered range, and it is well approximated by:
\begin{equation}\label{eq:Rcal}
\mathcal{R}(x,\delta)=\frac{x^{1+\epsRS(\dl)}(1-x)^{\btimplo(\dl)}}{\aRS(\dl)+\bRS(\dl) x+\cRS(\dl) x^2}\,,
\end{equation}
where $\btimplo(\dl)$ is given by Eq.~\ref{eq:Bguderlay}. 
The second term $\mathcal{F}(\om, \delta)$ is a re-scaling from the asymptotic curve to each $\om$ value and for all the $\dl$, with the form:
\begin{equation}\label{eq:Fcal}
\mathcal{F}(\omega,\delta)=1 + \frac{ a_F(\dl) \left[\Omega/\Omega_0(\dl) -1\right]}{ 1 +[\Omega/\Omega_0(\dl)]^{-2b_F(\dl)}}\,,
\end{equation}
where for writing simplicity we called: 
\begin{equation}
\Omega=1/(\om-5).
\end{equation}
All the parameters are given in Table~\ref{tab:PAR_RS}.

The combination of the two functions $\mathcal{R}(x,\delta)$ and $\mathcal{F}(\om,\delta)$, allows for having a smooth and continuous description of the RS curves in the given ranges of $\om$ and $\dl$. 
Anyway we notice that the function $\mathcal{R}(x,\delta)$ alone represents by itself a very good approximation for the RS curves for all the $\om\gtrsim8$, with an accuracy always better than 0.5\%.
%
\begin{table}
\small
\centering
\caption{Coefficients for RS global function.}
\label{tab:PAR_RS}
\begin{center}
\begin{tabular}{cc}
    \hline
    Function & Coefficient\\
    \hline
    \multirow{4}{*}{$\mathcal{R}$} & $\epsRS(\dl)=0.5548 + 0.03673 \,\dl$\\
    & $\aRS(\dl)=0.01964 - 0.01092 \,\dl$\\
    & $\bRS(\dl)=0.5095 - 0.09787\,\dl + 0.01412\,\dl^2$\\
    & $\cRS(\dl)=0.1871 + 0.1663 \,\dl$ \\
    \hline
    \multirow{3}{*}{$\mathcal{F}$} & $a_F(\dl)=0.02171 + 0.03051 \,\dl$\\
    & $b_F(\dl) = 1.389 - 0.3606 \,\dl$\\
    & $\Omega_0(\dl) = 0.3338 + 0.2884  \,\dl$ \\
    \hline
\end{tabular}
\end{center}
\end{table}
\begin{figure}
\centering
	\includegraphics[width=.5\textwidth]{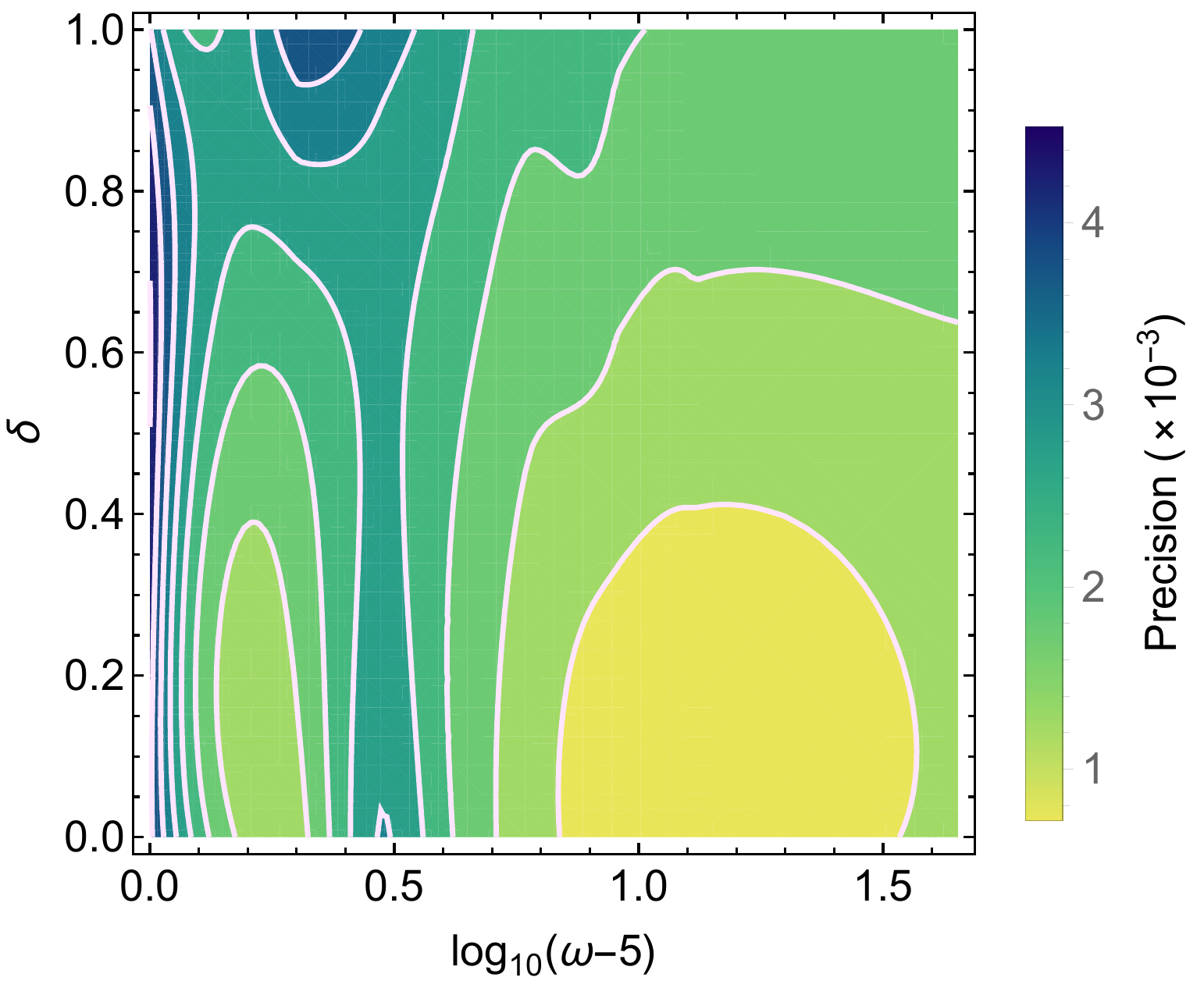}\,
    \caption{Contour map of the precision of the RS fit in the $\log_{10}(\om-5)--\dl$ plane.}
    \label{fig:RSFIT_accuracy}	
\end{figure}

In Fig.~\ref{fig:RSFIT_accuracy} we show the distribution of the errors introduced by our approximation to the RS evolution for each one of the cases we have investigated numerically. Errors are shown as contours in the ($\log_{10}(\om-5),\,\delta$) plane.
To summarize, we have then found a very good representation of the data, with our formulae being able to approximate all of our numerical profiles for the RS, in the parameters region $\om=\{6,50\}$ and $\dl=\{0,1\}$, with RMS errors always of order of a few $10^{-3}$.
We notice $\om=8$ to be the worst case in general (having a precision still better then $0.5\%$ for all $\dl$). This is not surprising, since it represents a transition point between the low-$\omega$ and high-$\omega$ regimes.
Finally, we note that deviations at lower $\om$ values from the shape given by Eq.~\ref{eq:Rcal} are more relevant only close to $\tcore$, while the behaviour near $\timplo$ stays always almost unaffected.

\subsection{Contact Discontinuity}

\begin{figure}
\centering
	\includegraphics[width=.49\textwidth]{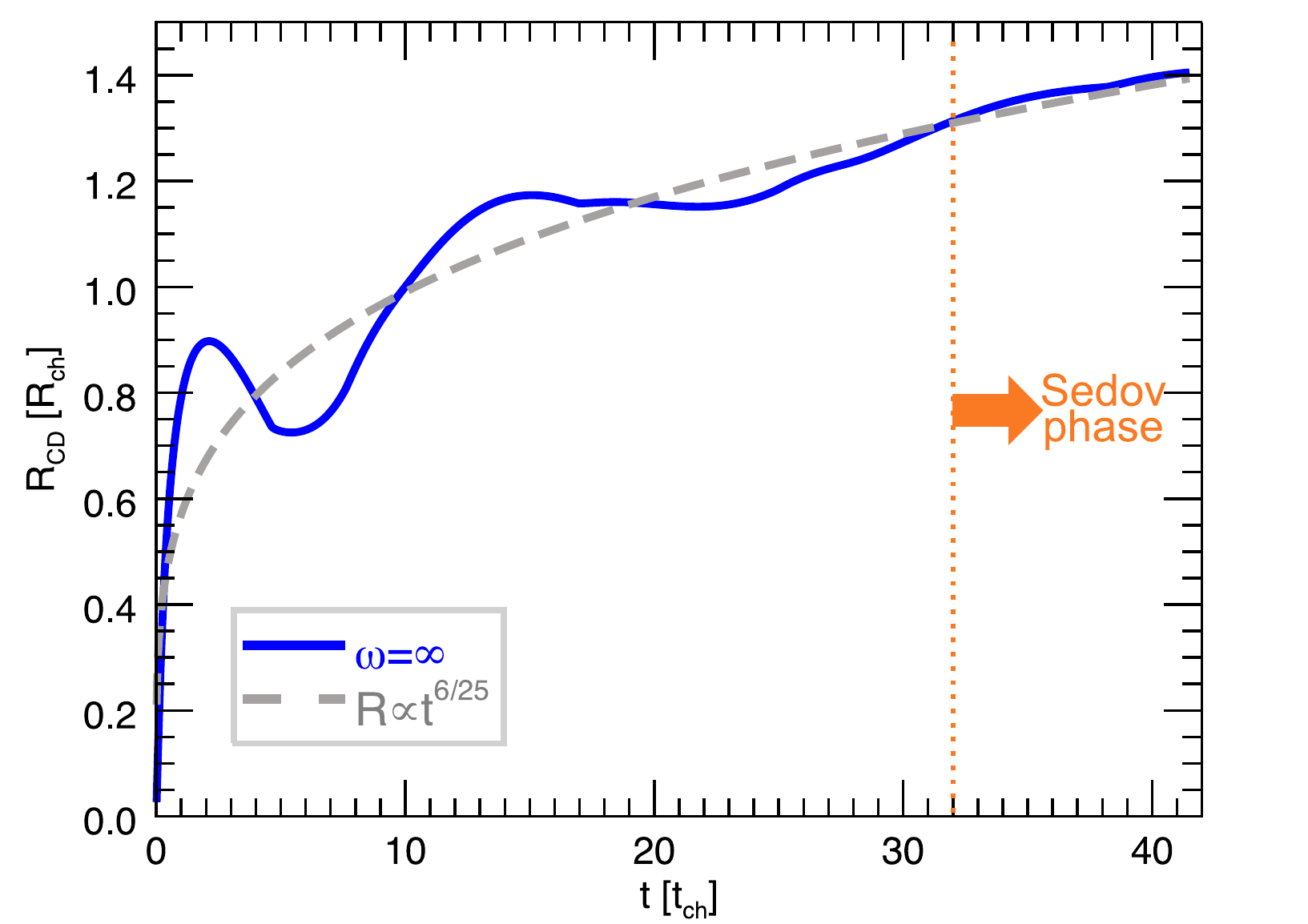}
    \caption{Long term evolution of the CD (for the $\om=\infty$ case) and comparison with the asymptotic trend, that signs the beginning of the Sedov phase.}
    \label{fig:CDlong}	
\end{figure}

The evolution of the CD, after $\tcoreCD$, shows a rather high level of complexity, characterized by oscillations with a damped amplitude.
The CD asymptotic evolution follows that of the FS: according to the Sedov solution, $\Rforw(t)\propto t^{2/5}$, while the pressure at the FS is $\sim\rho_0\Rforw(t)^2/t^2\propto t^{-6/5}$, and the pressure in the internal layers scales as $t^{-6/5}$ as well.
Since the medium inside the CD expands in an adiabatic way, asymptotically its pressure scales as $\Rcont^{-5}$, so that $\Rcont\propto t^{6/25}$. 
The above argument assumes an asymptotic pressure balance in the region surrounded by the CD, and we have verified that in the asymptotic regime it works rather well also quantitatively.
We did so by deriving --from extended-time numerical simulations-- an asymptotic value for the product of the internal energy inside the CD times $\Rcont^2$, and then assuming an homogeneous bubble, the asymptotic numerical evolution is well approximated.
Unfortunately, this asymptotic regime is reached only at very late times, beyond 30--40~$\tch$ (see Fig.~\ref{fig:CDlong}).
It is apparent that assuming the Sedov solution at times $\ll 30\,\tch$ is a strong enforcement, leading to an oversimplification of the problem and perhaps to a very different evolution of the system.

%

Before then, the radius of the CD experiences a number of prominent quasi-periodic oscillations, with a period increasing with time.
The reason of these oscillations can be qualitatively explained as the effect of reflected shock waves that bounce back and forth, both from the FS and from the RS; the period increase is likely associated with an increase of the sound crossing time.
The rather large elongation of these oscillations, in turn, can be justified by the fact that the density of the layers immediately surrounding the the CD is very low, and therefore in these conditions of low inertia even moderate pressure unbalances may cause rather large displacements.
Of course, such oscillations are also a numerical consequence of having assumed in our models a spherical symmetry; while in a real case this complex evolution is more likely to show a 3D rather chaotic behaviour \citep{Dwarkadas:2000}.

For the above reasons, in the following we will present accurate analytic approximations of the CD size evolution, but only valid at rather early times, namely not extending beyond $\timplo$.
A fit that includes the oscillatory phase is beyond our scope, and probably also of little use.

Even with this limitation, an accurate modelling of $\Rcont(t)$, valid over a wide range of $\om$ and $\dl$, is necessarily complex.
Among the various choices that we have tested, we have finally selected this one:
\begin{equation}
    \Rcont(t)\simeq\frac{\aCD(\om,\dl)t^{(\om-3)/\om}}{1+\bCD(\om,\dl)t^{\cCD(\om,\dl)}},
\end{equation}
where in the numerator we have kept the time dependence as in the C82 models.
The denominator, close to 1 as early times, is intended to rule how much the solution, beyond $\tcoreCD$, detaches from the extrapolation of the early self-similar evolution.
In principle, also the coefficient $\aCD(\om,\dl)$ could be derived from the C82 models but, since we have fitted these approximated formulae only to the numerical data beyond $\tcoreCD$, we have found slightly better fits not constraining this parameter.
The chosen functional dependencies are:
\begin{eqnarray}
    \aCD(\om,\dl)&=&a_{0,1}\dl+(1+a_{1,1}\dl)\,\tilde a(\om)\,, \\
    \bCD(\om,\dl)&=& b_{0,0}+b_{0,1}\dl+(b_{1,0}+b_{1,1}\dl)  \,\tilde a(\om)\,,    \\
    \cCD(\om,\dl)&=&\frac{c_{0,0}+c_{0,1}\dl+(c_{1,0}+c_{1,1}\dl)\,\om}{c_{2,0}+c_{2,1}\dl+\om}\,,
\end{eqnarray}
where:
\begin{equation}
\tilde a(\om)=\frac{A_0+B_0\,\om}{C_0+\om}\,,
\end{equation}
and all the best-fit parameters (15, all together), are listed in Table~\ref{tab:PAR_CD}.
%
\begin{table}
\small
\centering
\caption{Coefficients for CD fitting function.}
\label{tab:PAR_CD}
\begin{center}
\begin{tabular}{lc}
    \hline
    Function & Coefficient\\
    \hline
    $\aCD(\om,\dl)$& $a_{0,1}=-0.1597 \,, \;\; a_{1,1}=0.1859$\\
    \hline
    $\tilde a(\om)$ & $A_{0}=1.141\,, \;\; B_{0}=1.806\,, \;\; C_{0}=7.636$\\
    \hline
    \multirow{2}{*}{$\bCD(\om,\dl)$} & $b_{0,0}=-1.051,\,\;\; b_{0,1}=-0.1961$\\  
     & $b_{1,0}=1.290 \,,\;\; b_{1,1}=0.2375$\\
    \hline
    \multirow{3}{*}{$\cCD(\om,\dl)$} & $c_{0,0}=-5.561 \,,\;\; c_{0,1}=-0.6741$\\  
    & $c_{1,0}=1.265 \,,\;\; c_{1,1}=-0.07309$\\ 
    & $c_{2,0}=-4.826 \,,\;\; c_{2,1}=-0.6504$\\
    \hline
\end{tabular}
\end{center}
\end{table}

The accuracy of this approximation oscillates from 0.4\% to 0.8\%, for $\om$ values above 7; while it behaves not so well --in comparison-- for $\om$ in the range from 6 to 7, where the accuracy downgrades to a a few \% (with the worst case being $\om=6$ and $\dl=1$, for which we get a maximum error of 7\%).

\subsection{Forward Shock}

The case of the FS is slightly simpler than that of the CD for a few reasons: 
\begin{enumerate}
    \item the self-similar (C82) regime ends at a later time with respect to that of the CD;
    \item the analytic asymptotic solution (Sedov solution) is known, and it is equal to $\Rforw(t)=\xi_0 t^{2/5}$, with $\xi_0\simeq1.15169$ for $\Gamma=5/3$;
    \item the dependence on the density structure of the inner ejecta, namely on the parameters $\dl$ and $\om$, is expected to be weaker (see Fig.~\ref{fig:numerical_traj_allOM_D});
    \item finally, most of the swept-up mass is packed closer to the FS, and the inertia of those layers is higher than near to the CD, so that one does not expect the strong oscillations reported in the previous subsection.
\end{enumerate}
In this section we present a reasonably simple, but nonetheless accurate, analytic approximating function for the FS trajectory. 
We search in particular for the most appropriate way to join the asymptotic regime, in order to ensure the Sedov solution to be reached at late times.
Using as benchmark an extended-time simulation (in the specific case with an $\dl=0$, $\om=\infty$, and $t\rs{max}=14.4\,\tch$) we have found that the approximated function:
\begin{equation}
    \Rforw(t)=\frac{\xi_0\left(t+1.94\right)^{2/5}}{1+0.672/t+0.00373/t^2}\,\,
\end{equation}
where we recall that both the time and radii are expressed in characteristic units.
Even if there is a partial degeneracy among the best-fit parameters, the offset in time in the formula above, not only allows for a better accuracy, but can also be understood as the sign that the long-term effect of the earlier evolution, characterized by an expansion index $(\om-3)/\om$, higher than $2/5$, is an effective expansion time larger than the actual age.

The above formula allows to reproduce the FS trajectories for all $\om$ and $\dl$ values, with an accuracy always better than 2.5\%. 
Most of this error is actually accumulated close to $\tcoreFS$, while it reduces to a $\lesssim 1\%$ around  $\tch$, remaining of such order up to $\timplo$. 
We have also performed specific fits for all of our choices of the $\dl$ and $\om$ parameters, obtaining negligible differences in the parameters for each case.
%

\section{Discussion and Conclusions}
\label{sec:concl}
In this paper we have presented a detailed  model for the evolution and structure of nonradiative SNRs, in the one-dimensional spherically symmetric case.
Understanding this evolution, even in the considered simplified scheme, has been shown to be fundamental when modelling the complex dynamics of the interaction between the SNR and its host PWN \citep{Bandiera:2020}.
This interaction is very complex and a correct modelling, with high accuracy, of the SNR RS is particularly relevant. 
During their coupled evolution, the SNR and PWN start to interact directly during the so called reverberation phase, beginning when the RS, in its receding motion, encounters the PWN bubble.
Depending on the properties of the SNR at the beginning of reverberation and on the energetics of the PWN, the outcome of this interaction may substantially change the PWN structure.
The amount of compression the PWN will experience during this phase may determine all subsequent evolution, and generate local-in-time effects such as superefficiency \citep{TorresLin2018}.
Since it might produce important variations on the observational properties of the PWN, the reverberation phase is fundamental, and needs to be understood and correctly modeled. 
We recently started to investigate in detail the properties of the reverberation phase \cite[see][]{Bandiera:2020}, and our research program in this direction continues.
A deeper understanding of reverberation will especially affect the interpretation of aged systems --that will become more and more numerous in the near future,  thanks to the forthcoming new gamma-ray facilities (as the Cherenkov Telescope Array).
We found that even small variations of the position and velocity of the RS at the onset of reverberation may produce large variations in the final compression factor of the PWN (i.e. the ratio between the maximum and minimum radii the nebula experience during its evolution). This is already quite clear even by assuming the TM99 solutions on their own, and changing $\om$ only, as Fig. 2 of \citet{Bandiera:2020} shows. The effect is especially notable in the models presented in that paper, where we have considered the parameters of two sort of limit pulsars: a very energetic (the Crab) and a lower energetic (J1834.9-0846) ones.
We found that, in the former case, different $\om$ values lead to noticeable variations in the compression factor (from 3.5 to 11.4 in the range $0 \leq \om  \leq 12$). This effect is less evident for the lower energetic pulsar (even if the compression factor changes from 980 to $\sim$1400) but the time at which the maximum compression happens is still remarkably different.
We thus felt the need of developing a more accurate description of the RS structure and evolution; which is what we have presented here.

Using a Lagrangian numerical code, we have performed a large sample of numerical simulations and test their reliability comparing with the initial phase of the system evolution, for which analytical solutions are known (C82).
We have modeled the SNR ejecta considering a density profile with a radial power-law distribution both in the core (with index $\dl$) and in the envelope (with index $\om$) and repeated the simulations for a large set of different values of both parameters.

We have focused our investigation to the behaviour of the RS, but for the sake of completeness we have presented also analytic approximations for the evolution of the CD and the FS, with the hope that our results will have a wider range of applications, beyond our project on the  SNR-PWN interaction, started with \citet{Bandiera:2020} and that will continue with a forthcoming paper.

We found that the RS shows a fast monotonic convergence to the asymptotic case of the flat ejecta envelope (namely $\om=\infty$, or $\om=0$ in the TM99 notation) effect that is also not seen in the TM99 model.
Moreover our solutions for the RS evolution differ effectively from those of TM99, except for the asymptotic one and the $\om=6$ case, for which the difference is rather small.
However, deviations of at least 10\% are rather typical for a variety of $\om$ values and times.
Differences closer to the implosion time can easily reach values close to 100\% due to the loss of accuracy of the analytic formulas of TM99 in this final part of the evolution.
Beyond achieving a formal correction, these changes in the positions of the shock are indeed relevant: a variation of $\sim10\%$ in the RS position close to the beginning of reverberation  may change completely the outcome of this phase.
As already recalled previously, in \citet{Bandiera:2020} we have in fact shown that even the small variations of the RS profile, which occur by changing the $\om$ value within the TM99 model,  produce a different evolution of the PWN during the reverberation phase. 
We will discuss in a forthcoming paper how this picture changes taking into account our new model, for which we have shown relevant deviations from the TM99 one.

It is worth mentioning that our model is a simplification of the more complex 3-dimensional (3D) evolution of the SNR and its characteristic curves. 
A more realistic model would need to consider the results from 3D magneto-hydrodynamic simulations instead of 1D ones. 
Unfortunately such models are still not available for very long evolution or large parameters space; due to their huge computational cost, mainly models devoted to the description of specific systems have been produced, specialized to particular parameters of the SNR and ambient medium (see e.g. \citealt{Orlando:2019,Stockinger:2020,Tutone:2020,Gabler:2021} for recent results).
We expect that possible asymmetries introduced during the supernova explosion, as well as strong gradients in the ISM density, might change dramatically the evolution, since the spherical symmetry will be completely destroyed.
The development of strong instabilities driven at the boundaries (especially at the CD;  \citealt{Dwarkadas:2000}) can also introduce modifications to the spherical geometry.
On the other hand, we expect our model to be able to predict the position of the characteristic shocks for all the cases that would not differ much from the spherical case. As an example, \citet{Ferrand:2010} noticed a correspondence of the positions of characteristic shocks from their 3D simulations with the TM99 predictions.

We remark that the TM99 evolution of nonradiative supernova remnants has been used hundreds of time in the last 20 years, and affect aspects as varied as dust formation and survival in supernova ejecta \cite[see e.g.][]{Bianchi2007}. 
In this latter case, the passage of the RS  produces a shift of the size distribution function towards smaller grains of dust. 
Thus, a different velocity or position of the reverse and forward shock has an influence onto which grains, and of what size, can be formed in supernova remnants (see e.g. figure 7 of \citealt{Kozasa2009}).
We thus recommend a re-analysis of this issue using the solutions provided here.


\section*{Acknowledgements}
%
This work has been supported by grants PGC2018-095512-B-I00, SGR2017-1383, 2017-14-H.O ASI-INAF and INAF MAINSTREAM.
%

\appendix
\section{Notation}\label{app:notations}
In this Appendix we list the notation used within this paper and recall the one used in TM99 and C82, our reference works. 
For simplicity and consistency with our previous paper \citet{Bandiera:2020}, the used parameters name do not always coincide with those of TM99 and C82. 
We hope that the list in Table~\ref{tab:notation} can be of help when trying to migrate from one notation to another.
\begin{table}
\caption{List of the relevant parameters in our notation, compared with TM99 and/or C82 ones.}\label{tab:notation}
\scriptsize
\centering
\begin{tabular}{|c|c|c|c|}
  \hline
  Variable/Parameter & This Work & Value/Range  & Reference Work  \\
  \hline
  Ambient medium density & $\rho_0$ & --  & $\rho_0$
  (TM99), $q$ (C82) \\
  Ambient medium profile & --  & uniform & $s=0$ (TM99, C82) \\
  Ejecta envelope index & $\om$ & $>5$ &  $n$ (TM99, C82) \\
  Ejecta envelope limit case & $\om\rightarrow\infty$  & -- & $n=0$ (TM99) \\
  Envelope density parameter & $A v^{\om}_t$  & -- & $g_n$ (C82) \\
  Ejecta core index & $\dl$  & $0\leq \dl \leq 1$ & $0$ (TM99, C82) \\
  RS radius & $\Rrev$  & -- & $R^*_r$ (TM99), $R_1$ (C82) \\
  FS radius & $\Rforw$  & -- & $R^*_b$ (TM99), $R_2$ (C82) \\
  CD radius & $\Rcont$  & -- & $R_C$ (C82) \\

  \hline
\end{tabular}
\end{table}

%
\section{Table of the RS, CD and FS general formulas and direct comparison with TM99}\label{app:genTAB+cfrTM99}
In Table~\ref{tab:tabGEN} we show the complete set of formulas for the description of the RS, CD and FS evolution and their range of validity in terms of time and $\om$ values.
Our approximations have been obtained in the range $0\leq \dl \leq 1$.

In Table \ref{tab:TM99vsUS} we then compare our formulas for the characteristic curves, specialized to the $\dl=0$ case, directly with those from the TM99 paper. 
\begin{landscape}
\begin{table}
\caption{Summary of all the formulas approximating the characteristic curves for the RS, CD and FS evolution, with their relative validity ranges. \label{tab:tabGEN}} 
\fontsize{8.0}{7.8}\selectfont
\centering
        \makeatletter
           \def\rulecolor#1#{\CT@arc{#1}}
           \def\CT@arc#1#2{%
           \ifdim\baselineskip=\z@\noalign\fi
           {\gdef\CT@arc@{\color#1{#2}}}}
           \let\CT@arc@\relax
          \rulecolor{gray!50}
        \makeatother
\begin{tabular}{|l|c|c|c|c|c|c|c|c|c|c|}
  \hline
   \multicolumn{5}{|l|}{\bf{CH. CURVE}} & {\bf{FORMULAS}} & {\bf{VALIDITY}} &  {\bf{SUPPORTING DEFS.}}& {\bf{PARAMETERS}}  \\ 
   \multicolumn{5}{|l|}{} &  &  &   &  \\ 
\hline
 \multicolumn{5}{|l|}{RS} & $\mathcal{R}(x,\delta)\times \mathcal{F}(\om, \delta)$ & $\om\geq6\,,\;\tcore < t \leq \timplo$ & $\timplo=t^{\infty}_{\mathrm{implo}}  + \sqrt{ a_t^2\Omega^2 + \left[ (-b_t+c_t\Omega)/(1+1/\Omega^2)\right]^2}$  &  $t^{\infty}_{\mathrm{implo}}=2.399 + 0.4813\,\dl + 0.1760\,\dl^2$ \\ \multicolumn{5}{|l|}{} &  &  &   &   $a_t=0.1006 + 0.04184 \,\dl$ \\     
\multicolumn{5}{|l|}{} &  &  & $\Omega=1/(\om-5)$  &  $b_t= 0.06494 + 0.09363 \,\dl$  \\ 
\multicolumn{5}{|l|}{} &  &  &   &  $c_t=0.7063 -0.09444 \,\dl$  \\     
\multicolumn{5}{|l|}{} &  &  &   &       \\     
\multicolumn{5}{|l|}{} & $\mathcal{R}(x,\delta)=\left[x^{1+\epsRS}(1-x)^{\btimplo}\right]\left[\aRS+\bRS x+\cRS x^2\right]^{-1}$ &  &  & $\btimplo=0.68236+0.07720\,\dl+0.02456\,\dl^2$  \\
\multicolumn{5}{|l|}{} &  &  &   &  $\epsRS=0.5551 + 0.03674 \,\dl$\\   
\multicolumn{5}{|l|}{} &  &  &   &  $\aRS=0.01961 - 0.01091 \,\dl$\\ 
\multicolumn{5}{|l|}{} &  &  &   &  $\bRS=0.5093 - 0.09790\,\dl + 0.01412\,\dl^2$ \\ 
\multicolumn{5}{|l|}{} &  &  &   &  $\cRS=0.1874 + 0.1663 \,\dl$ \\ 
   \multicolumn{5}{|l|}{} & $\mathcal{F}(\omega,\delta)=1 + \bigg\{a_F \left[\Omega/\Omega_0 -1\right]\bigg\}\left\{ 1 +[\Omega/\Omega_0]^{-2b_F}\right\}^{-1}$ &  &  & $a_F=0.02171 + 0.03051 \,\dl$  \\   
\multicolumn{5}{|l|}{} &  &  &   &  $b_F = 1.389 - 0.3606 \,\dl$ \\  
\multicolumn{5}{|l|}{} &  &  &   &  $\Omega_0 = 0.3338 + 0.2884  \,\dl$ \\  
\multicolumn{5}{|l|}{} &  &  &  &   \\
\midrule
 \multicolumn{5}{|l|}{CD} & $\left[\aCD(\om,\dl)t^{(\om-3)/\om}\right]\left[1+\bCD(\om,\dl)t^{\cCD(\om,\dl)}\right]^{-1}$ & $\om\geq 6\,,\;\; \tcoreCD < t \leq \timplo$ & $\aCD=a_{0,1}\,\dl+(1+a_{1,1}\,\dl)\,\tilde a(\om)$ & $a_{0,1}=-0.1597\,,\;\; a_{1,1}=0.1859$\\  
 \multicolumn{5}{|l|}{} &  &  &  &   \\
 \multicolumn{5}{|l|}{} &  &  & $\bCD= b_{0,0}+b_{0,1}\,\dl+(b_{1,0}+b_{1,1}\,\dl)  \,\tilde a(\om)$ &  $b_{0,0}=-1.051\,,\;\; b_{0,1}=-0.1961$  \\
 \multicolumn{5}{|l|}{} &  &  &  & $b_{1,0}=1.290 \,,\;\; b_{1,1}=0.2375$  \\
  \multicolumn{5}{|l|}{} &  &  &  &   \\
 \multicolumn{5}{|l|}{} &  &  & $\cCD=\left[c_{0,0}+c_{0,1}\dl+(c_{1,0}+c_{1,1}\dl)\,\om\right]\left[c_{2,0}+c_{2,1}\dl+\om\right]^{-1} $ &  $c_{0,0}=-5.561 \,,\;\; c_{0,1}=-0.6741$ \\
 \multicolumn{5}{|l|}{} &  &  &  & $c_{1,0}=1.265 \,,\;\; c_{1,1}=-0.07309$  \\
 \multicolumn{5}{|l|}{} &  &  &  & $c_{2,0}=-4.826 \,,\;\; c_{2,1}=-0.6504$   \\
 \multicolumn{5}{|l|}{} &  &  &  &   \\
 \multicolumn{5}{|l|}{} &  &  & $\tilde a=\left(A_0+B_0\,\om\right)(C_0+\om)^{-1}$ & $A_{0}=1.141\,, \;\; B_{0}=1.806\,, \;\; C_{0}=7.636$  \\
 \multicolumn{5}{|l|}{} &  &  &  &   \\
 \midrule
 \multicolumn{5}{|l|}{FS} &  $\xi_0\left(t+1.94\right)^{2/5}\Big/\left[1+0.672/t+0.00373/t^2\right]$ & $\om\geq 6\,,\;\; t>\tcoreFS$ &  & $\xi_0=1.15169$ \\  
\multicolumn{5}{|l|}{} &  &  &  &   (from the standard Sedov solution)\\
 \hline
\multicolumn{1}{|l|}{The different times at which the characteristic curves enter the ejecta core are given in Sec.~\ref{sec:selfsim}, in particular see Eq.~\ref{eq:tcore},\ref{eq:tcoreCD} and \ref{eq:tcoreFS}.}\\
\end{tabular}
\end{table}
\end{landscape}
%
%
\begin{landscape}
\begin{table}
\caption{Direct comparison of our formulae for the characteristic trajectories RS, CD, FS  with those of TM99 (only valid for $\dl=0$). \label{tab:TM99vsUS}} 
\fontsize{8.0}{7.8}\selectfont
\centering
        \makeatletter
           \def\rulecolor#1#{\CT@arc{#1}}
           \def\CT@arc#1#2{%
           \ifdim\baselineskip=\z@\noalign\fi
           {\gdef\CT@arc@{\color#1{#2}}}}
           \let\CT@arc@\relax
          \rulecolor{gray!50}
        \makeatother
\begin{tabular}{|l|c|c|c|c|c|c|c|c|c|c|}
\hline
   \multicolumn{5}{|l|}{\multirow{3}{*}{{\bf{TM99}}}} &  &  &  & \\ 
   \multicolumn{5}{|l|}{} &  &  &   &  \\ 
   \multicolumn{5}{|l|}{} &  &   $\om$ & $t $ & SUPPORTING FORMULAS \\ 
\hline
   \multicolumn{5}{|l|}{} & $1.83t\left(1+3.26t^{3/2}\right)^{-2/3}$ & $\om=0$ & $t<t_{\rm{st}}$ &  $t_{\rm{st}}= \left( 2\om/[5(\om-3)] \hat{A}^{-\om/(\om-3)}\sqrt{2.026}\right)^{2\om/[3(\om-5)]}$ \\ 
   \multicolumn{5}{|l|}{} & $t(0.779 - 0.106 t - 0.533 \ln{t})$ & $ \om =0 $ & $t\geq t_{\rm{st}}$ & $\hat{A} = \left\{27 l_{\rm{ed}}^{(\om - 2)}/\left[4\upi\om(\om - 3)\phi_{\rm{ed}}\right]\left(10[\om - 5]/[3(\om - 3)]  \right)^{(\om - 
        3)/2}\right\}^{1/\om}$  \\ 
    \multicolumn{5}{|l|}{} &  &  &  & $l_{\rm{ed}}=\{1.39,\, 1.26,\, 1.21,\, 1.19,\, 1.17,\, 1.15,\, 1.14\}^*$  \\
    \multicolumn{5}{|l|}{} &  &  &  &   \\
    \multicolumn{5}{|l|}{} & $\hat{A} t^{(\om-3)/\om}$ & $5< \om \leq 14$ & $t\leq t_{\rm{core}}$ &  $t_{\rm{core}} = [27 / (4 \pi \om [\om - 3] l^2_{ed} \phi_{ed})]^{1/3} \sqrt{3 (\om - 3) / (10 [\om - 5])}$ \\ 
   \multicolumn{5}{|l|}{RS} &  &  &  &   \\
   \multicolumn{5}{|l|}{} & $t\left(\hat{A} t_{\rm{core}}^{(\om-3)/\om} \,\Big/\, l_{\rm{ed}}\right)$ & $5< \om \leq 14$ & $t_{\rm{core}}< t \leq t_{\rm{st}} $ & $\phi_{\rm{ed}}=\{0.39,\, 0.47,\, 0.52,\, 0.55,\, 0.57,\, 0.6,\, 0.62\}^*$ \\ 
   \multicolumn{5}{|l|}{} &  &  &  &   \\
   \multicolumn{5}{|l|}{} & $t \bigg[ (\hat{A} t_{\rm{core}}^{(\om-3)/\om})\,\Big/\, (l_{\rm{ed}} t_{\rm{core}}) - a_{\rm{core}} (t - t_{\rm{core}}) +$  &  &  &  \\ 
   \multicolumn{5}{|l|}{} & $-\left( 3\,\Big/\,( l_{\rm{ed}}\om) \hat{A} t_{\rm{core}}^{[(\om-3)/\om] - 1} - a_{\rm{core}} t_{\rm{core}}\right) \ln{(t/t_{\rm{core}})} \bigg]$ &  $5< \om \leq 14$ &  $ t > t_{\rm{st}} $ &  $a_{\rm{core}}=\{0.112,\, 0.116,\, 0.139,\, 0.162,\, 0.192,\, 0.251,\, 0.277\}^*$ \\
  \midrule
   \multicolumn{5}{|l|}{} &  &  &  &   \\
   \multicolumn{5}{|l|}{CD} & --- & ---  & --- & --- \\ 
   \multicolumn{5}{|l|}{} &  &  &  &   \\
   \midrule
  \multicolumn{5}{|l|}{} & $2.01t\left(1+1.72t^{3/2}\right)^{-2/3}$ & $\om=0$ & $t<t_{\rm{st}}$ &  \\ 
   \multicolumn{5}{|l|}{} & $(1.42 t - 0.254)^{2/5}$ & $ \om =0 $ & $t\geq t_{\rm{st}}$ &   \\ 
   \multicolumn{5}{|l|}{FS} &  &  &  &   \\
   \multicolumn{5}{|l|}{} & $\hat{A}t^{(\om-3)/\om}$ &  $5< \om \leq 14$ & $t \leq t_{\rm{st}}$ &   \\ 
   \multicolumn{5}{|l|}{} &    $\left(\hat{A}t_{\rm{st}}^{(\om-3)/\om}\right)^{5/2} + \sqrt{2.026}\left(t-t_{\rm{st}}\right)^{2/5}$ &  $5< \om < 14$ & $t > t_{\rm{st}}$ &   \\
  \hline
     \multicolumn{1}{|l|}{*All the listed numerical values must be considered relative to the range
   $\om=\{6,\,7,\,8,\,9,\,10,\,12,\,14\}$.}\\
        \multicolumn{1}{|l|}{  }\\
   \multicolumn{5}{|l|}{\multirow{3}{*}{\bf{THIS}}} &  &  &  & \\ 
   \multicolumn{5}{|l|}{} &  &  &   &  \\ 
   \multicolumn{5}{|l|}{} {\hspace{-1.5cm}\bf{WORK}} &  &   $\om$ & $t$ & SUPPORTING FORMULAS \\ 
\hline
 \multicolumn{5}{|l|}{} & &  &  \\  
   \multicolumn{5}{|l|}{} &  &  &  & $x=t/\timplo(\om,\dl)$ \\
   \multicolumn{5}{|l|}{} &  &  &  &  \\
   \multicolumn{5}{|l|}{} &  &  &  & $\timplo= 2.399 + \Bigg\{ (0.1006\Omega)^2  + $\\
  \multicolumn{5}{|l|}{}  &  &  &  &  $ + \left[ (-0.06494+0.7063\Omega)\Big/(1+1/\Omega^2)\right]^2
  \Bigg\}^{0.5} $\\
 \multicolumn{5}{|l|}{RS} & $\mathcal{R}(x)\times \mathcal{F}(\om)$ &  $\om\geq6$  & $\tcore < t \leq \timplo$   &  \\
 \multicolumn{5}{|l|}{} &  &  &   & $\Omega=1/(\om-5)$ \\ 
  \multicolumn{5}{|l|}{} &  &  &   &  \\
   \multicolumn{5}{|l|}{} &  &   &   &   $\mathcal{R}(x)=\left[x^{1.5551}(1-x)^{0.68236}\right]\left[0.01961+0.5093 x+0.1874  x^2\right]^{-1}$\\
 \multicolumn{5}{|l|}{} &  &  &   &  \\
   \multicolumn{5}{|l|}{} &  &  &  &  $\mathcal{F}(\omega)=1 + \bigg\{0.02171 \left[\Omega/0.3338 -1\right]\bigg\}\left\{ 1 +[\Omega/0.3338]^{-2.778}\right\}^{-1}$\\
    \multicolumn{5}{|l|}{} &  &  &   &  \\
\midrule
 \multicolumn{5}{|l|}{\multirow{10}{*}{CD}} & &  &  \\ 
   \multicolumn{5}{|l|}{} & $ \tilde a(\om)t^{(\om-3)/\om}    \Bigg/\,\left[1+\bCD(\om)t^{\cCD(\om)}\right] $ & $\om\geq6$ & $\tcoreCD < t \leq \timplo$ & $\tilde a(\om)=\left(1.141+1.806\,\om\right)(7.636+\om)^{-1} $ \\
   \multicolumn{5}{|l|}{} &  &  &  &   \\
   \multicolumn{5}{|l|}{} & &  &   & $\bCD=-1.051-0.1961\, \tilde a(\om)$  \\
     \multicolumn{5}{|l|}{} &  &  &  &   \\
  \multicolumn{5}{|l|}{} &  &  &  & $\cCD=-\left(5.561+0.6741\,\om\right) \,\left(\om -4.826+ \right)^{-1}$ \\

\midrule
 \multicolumn{5}{|l|}{} & &  &  \\ 
   \multicolumn{5}{|l|}{FS} & $\xi_0\left(t+1.94\right)^{2/5}\Bigg/\left[1+0.672(1/t)+0.00373(1/t)^2\right]$ & $\om\geq 6$ & $t>\tcoreFS$ & $\xi_0=1.15169$  \\
      \multicolumn{5}{|l|}{} &  &  &  &  (from the standard Sedov solution) \\
   \multicolumn{5}{|l|}{} &  &  &  &   \\
 \hline
\end{tabular}
\end{table}
\end{landscape}

%
%

\section*{Data availability}
The data underlying this article are available in the article.

\bibliographystyle{mn2e} 
\bibliography{biblio}

\label{lastpage}
\end{document}